\documentclass[aps,prb,twocolumn,preprintnumbers,amsmath,amssymb,superscriptaddress]{revtex4}%

\usepackage{graphicx}
\usepackage{dcolumn}
\usepackage{amsmath}
\usepackage{color}
\usepackage{hyperref}
\usepackage{bbold}
\usepackage{multirow}

\newcommand{\RN}[1]{\textup{\uppercase\expandafter{\romannumeral#1}}}%

\begin{document}

\title{Effect of Pressure and Oxygen-Isotope Substitution on Density-Wave Transitions in La$_4$Ni$_3$O$_{10}$}

\author{Rustem Khasanov}
 \email{rustem.khasanov@psi.ch}
 \affiliation{PSI Center for Neutron and Muon Sciences CNM, 5232 Villigen PSI, Switzerland}

\author{Vahid Sazgari}
\affiliation{PSI Center for Neutron and Muon Sciences CNM, 5232 Villigen PSI, Switzerland}

\author{Thomas J. Hicken}
% \email{thomas.hicken@psi.ch}
 \affiliation{PSI Center for Neutron and Muon Sciences CNM, 5232 Villigen PSI, Switzerland}

\author{Igor Plokhikh}
% \email{igor.plokhikh@psi.ch; i.plokhikh@fz-juelich.de}
 \affiliation{PSI Center for Neutron and Muon Sciences CNM, 5232 Villigen PSI, Switzerland}
 \affiliation{TU Dortmund University, Department of Physics, Dortmund, 44227, Germany}

\author{Marisa Medarde}
 \affiliation{PSI Center for Neutron and Muon Sciences CNM, 5232 Villigen PSI, Switzerland}

\author{Ekaterina Pomjakushina}
 \affiliation{PSI Center for Neutron and Muon Sciences CNM, 5232 Villigen PSI, Switzerland}

\author{Lukas Keller}
 \affiliation{PSI Center for Neutron and Muon Sciences CNM, 5232 Villigen PSI, Switzerland}

\author{Vladimir Pomjakushin}
 \affiliation{PSI Center for Neutron and Muon Sciences CNM, 5232 Villigen PSI, Switzerland}

\author{Marek Bartkowiak}
 \affiliation{PSI Center for Neutron and Muon Sciences CNM, 5232 Villigen PSI, Switzerland}

\author{Szymon Kr\'{o}lak}
 \affiliation{Faculty of Applied Physics and Mathematics, Gda\'{n}sk University of Technology, Narutowicza 11/12, Gdansk, 80-233 Poland}
 \affiliation{Advanced Materials Center, Gda\'{n}sk University of Technology, Narutowicza 11/12, Gdansk, 80-233 Poland}

\author{Micha{\l} J. Winiarski}
 \affiliation{Faculty of Applied Physics and Mathematics, Gda\'{n}sk University of Technology, Narutowicza 11/12, Gdansk, 80-233 Poland}
 \affiliation{Advanced Materials Center, Gda\'{n}sk University of Technology, Narutowicza 11/12, Gdansk, 80-233 Poland}

\author{Alexander Steppke}
 \affiliation{PSI Center for Neutron and Muon Sciences CNM, 5232 Villigen PSI, Switzerland}

\author{Jonas A. Krieger}
 \affiliation{PSI Center for Neutron and Muon Sciences CNM, 5232 Villigen PSI, Switzerland}

\author{Hubertus Luetkens}
 \affiliation{PSI Center for Neutron and Muon Sciences CNM, 5232 Villigen PSI, Switzerland}

\author{Tomasz Klimczuk}
 \affiliation{Faculty of Applied Physics and Mathematics, Gda\'{n}sk University of Technology, Narutowicza 11/12, Gdansk, 80-233 Poland}
 \affiliation{Advanced Materials Center, Gda\'{n}sk University of Technology, Narutowicza 11/12, Gdansk, 80-233 Poland}

\author{Christof W. Schneider}
 \affiliation{PSI Center for Neutron and Muon Sciences CNM, 5232 Villigen PSI, Switzerland}

\author{Dariusz J. Gawryluk}
 \affiliation{PSI Center for Neutron and Muon Sciences CNM, 5232 Villigen PSI, Switzerland}

\author{Zurab Guguchia}
 \affiliation{PSI Center for Neutron and Muon Sciences CNM, 5232 Villigen PSI, Switzerland}

\begin{abstract}
Understanding the interplay between magnetism and superconductivity in nickelate systems is a key objective in condensed matter physics. Gaining microscopic insights into magnetism -- particularly as it emerges near superconductivity -- requires a synergistic approach that combines complementary experimental techniques with controlled tuning of external parameters.
In this paper, we present a systematic investigation of the three-layer Ruddlesden–Popper (RP) nickelate La$_4$Ni$_3$O$_{10}$ using muon-spin rotation/relaxation ($\mu$SR) and resistivity measurements.
At ambient pressure, two incommensurate spin-density wave (SDW) transitions are identified at $T_{\rm SDW} \simeq 132$~K and $T^\ast \simeq 80-90$~K. Comparison of the observed internal magnetic fields with dipole-field calculations reveals a magnetic structure consistent with antiferromagnetically coupled SDW order on the outer two Ni layers, with smaller moments on the inner Ni layer. Above $T^\ast$, the moments lie primarily in the $ab$-plane, but below this temperature they undergo a subtle distortion and develop a $c$-axis component.
The internal fields at the muon stopping sites appear abruptly at $T_{\rm SDW}$, suggesting a first-order-like nature of the SDW transition, which is closely linked to the charge-density wave (CDW) order occurring at the same temperature ($T_{\rm SDW} = T_{\rm CDW}$).
Under applied pressure, all transition temperatures -- including $T_{\rm SDW}$, $T^\ast$, and $T_{\rm CDW}$ -- are suppressed at a nearly uniform rate of $\simeq -13$~K/GPa. This behavior contrasts with that of the two-layer RP nickelate La$_3$Ni$_2$O$_7$, where pressure enhances the separation between the SDW and CDW transitions.
The oxygen-isotope substitution ($^{16}$O $\rightarrow$ $^{18}$O) reveals that the CDW transition temperature shifts to higher values in the $^{18}$O-substituted samples. The isotope effect on $T_{\rm SDW}$ and $T^\ast$ differs significantly. Specifically, when the CDW and SDW orders are intertwined, a notable isotope effect is observed on $T_{\rm SDW}$, leading to equal transition temperatures and nearly identical isotope shifts for both $T_{\rm CDW}$ and $T_{\rm SDW}$. In contrast, at $T^\ast$, where the SDW transition occurs independently of the CDW, no isotope effect is detected.
\end{abstract}

%\pacs{74.70.Xa, 74.25.Bt, 74.45.+c, 76.75.+i}
\maketitle

%Introduction

\section{Introduction}

The exploration of nickelates, including those belonging to the Ruddlesden-Popper (RP) series La$_{(n+1)-x}$Pr$_x$Ni$_n$O$_{3n+1}$ ($n=2,3$), has garnered significant attention in condensed matter physics and materials science due to their intriguing structural, electronic, and magnetic properties\cite{Sun_Nature_2023, Zhang_NatCom_2020, Zhu_Nature_2024, Sakakibara_PRB_2024, Wang_CPL_2024, Zhang_arxiv_2023, Li_CPL_2024, Xu_arxiv_2025, Wu_PRB_2001, Huangfu_PRR_2020, Li_NatCom_2017, Zhang_PRM_2020, Khasanov_La327_arxiv_2024, Wang_Nature_2024, Pei_arxiv_2024, Shi_arxiv_2025, Li_arxiv_2025, Zhang_arxiv_2025,  Huangfu_PRB_2020, Huangfu_PRB_2020_2, Carvalho_JAP_2000, Seo_InorgChem_1966, Li_SCiChina_2024}. The recent observation of high-temperature superconductivity under high pressure in a member of the RP series with $n = 2$, has further enhanced interest in this class of materials, highlighting their potential for groundbreaking discoveries. Relevant studies include recent works on La$_{3-x}$Pr$_x$Ni$_2$O$_7$ under high pressure, which demonstrate its superconducting properties,\cite{Sun_Nature_2023, Wang_CPL_2024, Wang_Nature_2024}  as well as investigations into the electronic and structural properties of La$_{4-x}$Pr$_x$Ni$_3$O$_{10}$\cite{Zhang_NatCom_2020, Wu_PRB_2001, Wang_Nature_2024, Pei_arxiv_2024, Zhang_arxiv_2025, Huangfu_PRB_2020, Huangfu_PRB_2020_2}.  Among these, the compound La$_4$Ni$_3$O$_{10}$, a member of the RP series with $n = 3$, stands out due to its proximity to potential metal-to-metal transitions\cite{Huangfu_PRR_2020, Li_NatCom_2017, Zhang_PRM_2020, Carvalho_JAP_2000, Seo_InorgChem_1966}, charge ordering\cite{Zhang_NatCom_2020, Wu_PRB_2001}, and unconventional superconductivity\cite{Zhu_Nature_2024, Sakakibara_PRB_2024, Zhang_arxiv_2023}. Such features make it a promising candidate for understanding correlated electron systems and for exploring novel phenomena under extreme conditions.

%It should also be noted that resistivity curves for La$_3$Ni$_2$O$_7$ -- at least those measured under pressure -- do not clearly indicate an SDW transition, but do exhibit features consistent with a CDW transition.\cite{Sun_Nature_2023, Wang_CPL_2024, Wu_PRB_2001, Khasanov_La327_arxiv_2024}

At ambient pressure, La$_4$Ni$_3$O$_{10}$ exhibits two types of order: charge density wave (CDW) and spin density wave (SDW) states. These states have similar ordering temperatures and are strongly intertwined, highlighting the complex interplay between charge and spin degrees of freedom in this material~\cite{Zhang_NatCom_2020}.
The phase diagram of La$_4$Ni$_3$O$_{10}$ reveals a delicate balance between the density wave (DW) states and the superconducting phase, with pressure acting as a key tuning parameter~\cite{Zhu_Nature_2024, Sakakibara_PRB_2024, Zhang_arxiv_2023}. As pressure increases, the suppression of DW order appears to coincide with the emergence of superconductivity, indicative of a strong interplay between these competing phases.
However, the exact nature of the DW state suppressed by external pressure remains unidentified. For comparison, in the double-layer RP nickelate La$_3$Ni$_2$O$_7$, the SDW and CDW ordering temperatures ($T_{\rm SDW}$ and $T_{\rm CDW}$) exhibit opposite trends: $T_{\rm SDW}$, as defined by $\mu$SR experiments, increases with pressure, while $T_{\rm CDW}$, determined from resistivity measurements, decreases~\cite{Khasanov_La327_arxiv_2024}.

Beyond pressure, isotope substitution was shown to influence the charge-density wave (CDW) and spin-density wave (SDW) transition temperatures in various materials. \cite{Khasanov_PRL_YPa123_2008, Guguchia_PRL_2014, Medarde_PRL_1998, Luetkens_JMMM_2007, Amit_AdvCondMat_2011, Shengelaya_PRL_1999, Zhao_PRB_1994, Lanzara_JPCM_1999, Guguchia_PRB_2015, Bendele_PRB_2017} The formation of charge and spin-density waves arises due to collective electronic instabilities, often accompanied by lattice distortions, making these phases highly sensitive to phononic contributions. Variations in the transition temperatures ($T_{\text{CDW}}$ and $T_{\text{SDW}}$) upon isotope substitution provide insight into the relative importance of electron-phonon interactions versus purely electronic effects in stabilizing these ordered states.

This paper presents a systematic investigation of La$_4$Ni$_3$O$_{10}$ using a combination of muon-spin rotation/relaxation ($\mu$SR) and resistivity measurements. These complementary techniques provide detailed insights into the magnetic and electronic behavior of the material under various conditions -- particularly at ambient and high pressures, as well as upon substitution of $^{16}$O with the $^{18}$O isotope. The main findings are summarized as follows:
\begin{itemize}
  \item At ambient pressure, La$_4$Ni$_3$O$_{10}$ exhibits two spin-density wave (SDW) transitions at $T_{\rm SDW} \simeq 132$~K and $T^\ast \simeq 80-90$~K, with magnetic moments preferentially aligned within the $ab$-planes. Comparison of the observed internal magnetic fields with dipolar-field calculations reveals a magnetic structure consistent with antiferromagnetically coupled SDW order on the outer two Ni layers and smaller moments on the inner Ni layer. The transition at $T^\ast$ is further associated with a subtle reorientation of Ni moments, acquiring a $c$-axis component at lower temperatures.

  \item Under applied pressure, all transition temperatures — $T_{\rm SDW}$, $T^\ast$, and $T_{\rm CDW}$ — are suppressed. Despite this, the SDW and CDW transitions remain closely linked and are most likely strongly intertwined. This behavior contrasts with that of the two-layer Ruddlesden–Popper (RP) nickelate La$_3$Ni$_2$O$_7$, where pressure induces an increasing separation between the two density-wave orders~\cite{Khasanov_La327_arxiv_2024}.

  \item Oxygen-isotope effect (OIE) experiments reveal that when the CDW and SDW transitions are intertwined ($T_{\rm CDW} \simeq T_{\rm SDW}$), the ordering temperatures shift to higher values in the $^{18}$O-substituted sample. The isotope shifts are found to be comparable for both transitions, with $^{18}T_{\rm CDW}-^{16}T_{\rm CDW}\simeq^{18}T_{\rm SDW}-^{16}T_{\rm SDW}\simeq2.1$~K. In contrast, the OIE on the spin-reorientation transition at $T^\ast$ -- where the SDW state evolves independently from the CDW -- is negligible.
\end{itemize}
\noindent
Our results underscore the intertwined nature of SDW and CDW order in La$_4$Ni$_3$O$_{10}$, consistent with behavior observed in other correlated systems such as the cuprates~\cite{Tranquada_Nature_1995, Ghiringhelli_Sciense_2012, Chang_NatPhys_2012, Fradkin_RMP_2015, Keimer_Nature_2015, Kivelson_RMP_2003, Guguchia_PRL_2020} and hole-doped nickelates~\cite{Ricci_PRL_2021, Zhang_PNAS_2016, Zhang_PRL_2019}.
Given the distinct pressure-dependent evolution of SDW order observed in La$_3$Ni$_2$O$_7$ (Ref.~\onlinecite{Khasanov_La327_arxiv_2024}) and La$_4$Ni$_3$O$_{10}$ (this work), we propose that -- if the behavior of the density wave transitions persists at higher pressures -- the pressure-induced suppression of CDW order may be a possible mechanism driving the emergence of high-temperature superconductivity in Ruddlesden–Popper nickelates. Notably, while $T_{\rm CDW}$ decreases with pressure in both systems, $T_{\rm SDW}$ evolves differently: it increases in La$_3$Ni$_2$O$_7$ but decreases in La$_4$Ni$_3$O$_{10}$, underscoring fundamental differences in their electronic phase behavior.

\begin{figure*}[htb]
\includegraphics[width=1.0\linewidth]{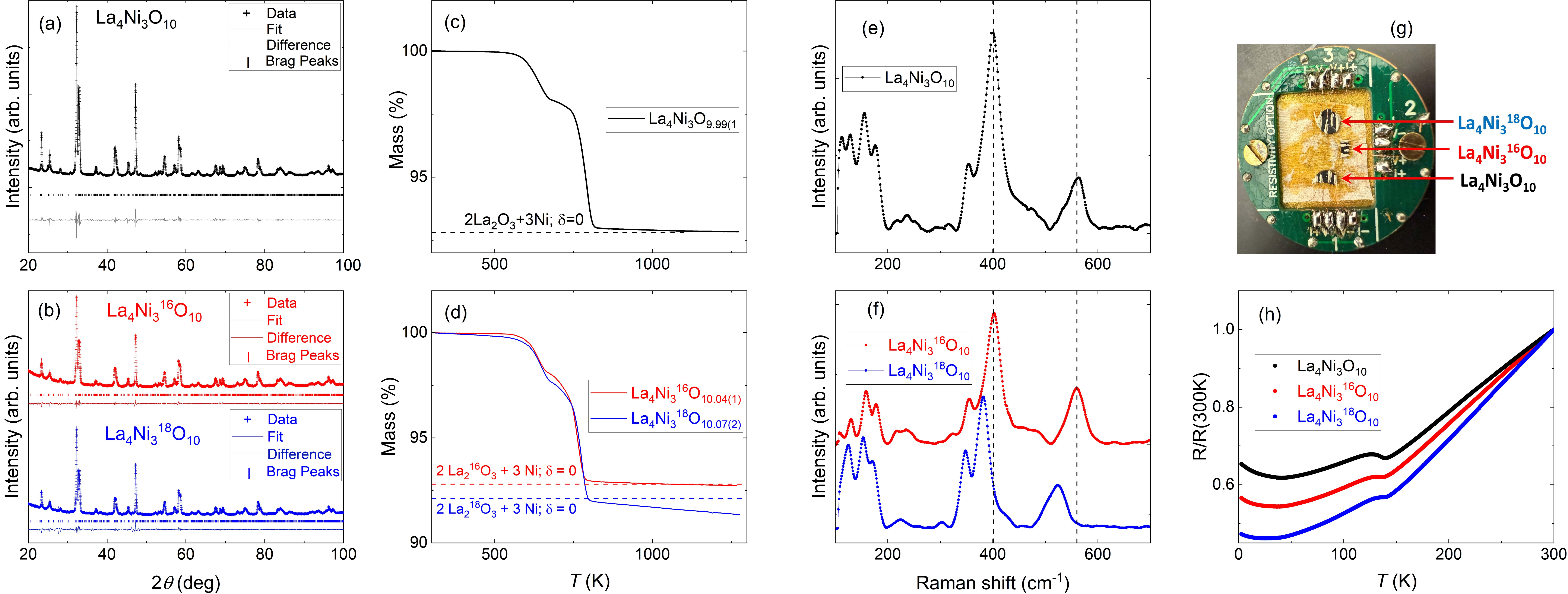}
\caption{
%{\bf Characterization of La$_4$Ni$_3$O$_{10}$ samples.}
(a)--(b) X-ray diffraction patterns of pristine [panel (a)] and oxygen-isotope ($^{16}$O/$^{18}$O) substituted [panel (b)] La$_4$Ni$_3$O$_{10}$ samples taken at room temperature. The x-ray data were refined using the $P2_1/a$ [panel (a)] and $Bmab$ [panel (b)] structures.
(c)--(d) Thermogravimetric curves of the pristine [panel (c)] and oxygen-isotope substituted [panel (d)] La$_4$Ni$_3$O$_{10}$ samples.
(e)--(f) Room-temperature Raman spectra collected for pristine [panel (e)] and oxygen-isotope-substituted [panel (f)] La$_4$Ni$_3$O$_{10}$ samples. The dashed lines indicate several Raman modes of the samples containing the lighter oxygen isotope.
(g) Photograph of polycrystalline La$_4$Ni$_3$O$_{10}$, La$_4$Ni$_3\,^{16}$O$_{10}$, and La$_4$Ni$_3\,^{18}$O$_{10}$ samples mounted on the measurement chip and prepared for resistivity measurements.
(h) Temperature dependence of resistivity normalized to its 300 K value $R(T)/R(300)$ for pristine and isotope-substituted La$_4$Ni$_3$O$_{10}$ samples.
} \label{fig:Characterization}
\end{figure*}

The paper is organized as follows. Section~\ref{sec:characterization} describes the characterization of La$_4$Ni$_3$O$_{10-\delta}$ samples using x-ray diffraction, thermogravimetry, neutron powder diffraction, and resistivity measurements. Section~\ref{sec:ambient-pressure} presents zero-field $\mu$SR measurements on the pristine La$_4$Ni$_3$O$_{10-\delta}$ sample, along with results of muon-stopping site and dipolar field calculations. Experiments conducted under hydrostatic pressure conditions are detailed in Section~\ref{sec:pressure-experiments}. Section~\ref{sec:OIE} presents Raman, $\mu$SR and resistivity studies of $^{16}$O/$^{18}$O isotope-substituted samples. The conclusions follow in Section~\ref{sec:Conclusions}.

\section{Sample characterisation} \label{sec:characterization}

The initially grown La$_4$Ni$_3$O$_{10-\delta}$ sample was divided into three parts. The first part, hereafter referred to as the "pristine" sample, was used for ambient- and high-pressure experiments. The remaining two parts were used to prepare oxygen-isotope-substituted samples (La$_4$Ni$_3\,^{16}$O$_{10}$ and La$_4$Ni$_3\,^{18}$O$_{10}$) by annealing them in either a $^{16}$O$_2$ or $^{18}$O$_2$ atmosphere.

Figures~\ref{fig:Characterization}(a) and (b) show the x-ray diffraction patterns collected at room temperature. Rietveld refinement of the x-ray data for the pristine sample yields a monoclinic structure ($P2_1/a$, space group No.~14),  with lattice parameters $a = 5.4162(3)$~\AA, $b = 5.4642(2)$~\AA, $c = 27.984(1)$~\AA, and $\beta = 90.256(4)^\circ$,
in agreement with previous reports~\cite{Zhang_NatCom_2020, Zhu_Nature_2024, Sakakibara_PRB_2024, Zhang_arxiv_2023, Li_SCiChina_2024}.
To reduce the number of refined parameters for the oxygen-isotope substituted-samples  -- and noting that the monoclinic $P2_1/a$ symmetry with $\beta \simeq 90^\circ$ is only marginally distorted from the orthorhombic $Bmab$ space group -- the x-ray data for all La$_4$Ni$_3$O$_{10-\delta}$ samples were further refined using the $Bmab$ symmetry.
The results of the x-ray refinements are summarized in Table~\ref{table:sample-characterzation}. All samples exhibit nearly identical lattice constants.

\begin{table}[htb]
\caption{Results of characterisation of La$_4$Ni$_3$O$_{10}$ samples by means of x-ray and thermogravimetry.}
\begin{tabular}{c|c|c|ccccc}
%  \hline
    &La$_4$Ni$_3$O$_{10-\delta}$&La$_4$Ni$_3\, ^{16}$O$_{10-\delta}$&La$_4$Ni$_3\, ^{18}$O$_{10-\delta}$&  \\
  \hline
Group symmetry& $Bmab$       &$Bmab$     &$Bmab$\\
$a$~(\AA) & 5.4140(3)  & 5.4163(3)   & 5.4134(3)\\
$b$~(\AA) & 5.4622(2) & 5.4623(3)   & 5.4602(2)\\
$c$~(\AA) & 27.9766(1)  & 27.9710(1)   & 27.9715(3)\\
\hline
Oxygen Content&  9.99(1)    &  10.04(1)   &10.04(2) \\
%\hline
\end{tabular}
\label{table:sample-characterzation} % is used to refer this table in the text
\end{table}

The oxygen content ($10-\delta$) was determined using thermogravimetric analysis, as shown in Figs.~\ref{fig:Characterization}~(c) and (d). A small amount of material ($\simeq 20-40$~mg) was heated from room temperature to 1000$^{\circ}$C at a rate of 1$^{\circ}$C/min in a flowing H$_2$/He gas mixture (5 vol.\% hydrogen and 95 vol.\% helium). The total weight loss was attributed to the reduction of fully oxidized samples to metallic nickel and La$_2$O$_3$.
Two important points need to be mentioned:\\
(i) The weight loss in the $^{18}{\rm O}$-substituted samples was bigger than that in the pristine and $^{16}{\rm O}$-substituted ones due to the heavier mass of $^{18}{\rm O}$ atoms compared to $^{16}{\rm O}$.\\
(ii) Above $T \sim 500^{\circ}$C, the decomposed $^{18}{\rm O}$-substituted sample continue to lose weight, whereas the weight remains nearly constant for the pristine and $^{16}{\rm O}$-substituted ones. This behavior may be attributed to the partial exchange of $^{18}{\rm O}$ atoms in La$_2$O$_3$ with residual $^{16}{\rm O}$ from the H$_2$/He gas mixture. \\
The results of the thermogravimetric analysis are summarized in Table~\ref{table:sample-characterzation}. Considering the (ii) point above, the La$_4$Ni$_3$O$_{10-\delta}$ samples studied here can be considered nearly oxygen-stoichiometric.

The room temperature Raman experiments confirm the appearance of several lines in the region ranging from 100 up 700~cm$^{-1}$ in agreement with the results of Refs.~\onlinecite{Li_SciBul_2025, Gim_arxiv_2025, Deswal-arxiv_2025, Suthar_arxiv_2025}. The Raman lines stay nearly at the same positions at pristine and $^{16}{\rm O}$-substituted samples, while they are clearly shifted to the lower frequencies in $^{18}{\rm O}$-substituted ones, see Figs~\ref{fig:Characterization}~(e) and (f). In order to ensure on the uniform distribution of oxygen isotopes, several experiments at different laser beam positions were performed.

The resistivity measurements were performed simultaneously on the three samples, all mounted on the same measurement chip [see Fig.~\ref{fig:Characterization}~(g)]. The resulting resistivity [$R(T)$] curves are shown in Fig.~\ref{fig:Characterization}~(h). The $R(T)$ dependencies exhibit a weak anomaly near $T \simeq 130$~K, which is attributed in the literature to the onset of intertwined spin-density wave (SDW) and charge-density wave (CDW) orders~\cite{Zhang_NatCom_2020, Zhu_Nature_2024}.

High-resolution neutron powder diffraction (NPD) data were collected at $T = 1.7$, 100, and 150~K using the HRPT diffractometer. The results indicate the absence of a structural phase transition within this temperature range. The refined crystal structure is consistent with the $P2_1/a$ symmetry previously determined from x-ray diffraction \cite{deposition number}. The temperature evolution of the incommensurate spin-density wave (SDW) magnetic peak at $(0, 0.574, 2)$ was monitored by collecting high-intensity NPD patterns using the DMC diffractometer. The results of the neutron powder diffraction studies on pristine La$_4$Ni$_3$O$_{10}$ samples are presented in the Supplemental Material \cite{Supplementa_Information}.

\begin{figure*}[htb]
\includegraphics[width=1\linewidth]{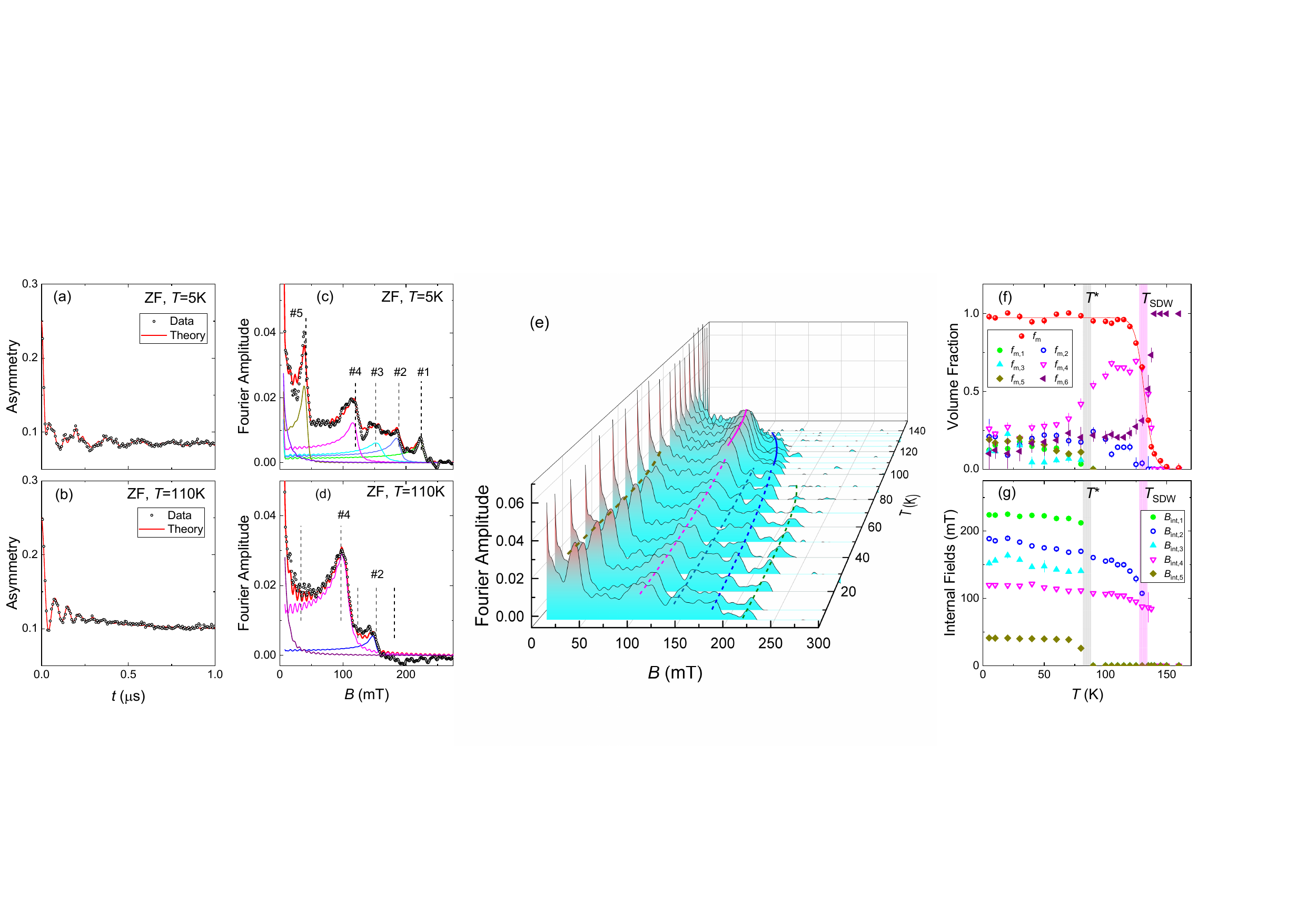}
\caption{
%{\bf Results of ambient-pressure ZF-$\mu$SR experiments.}
(a)--(b) Zero-field $\mu$SR time spectra of the pristine La$_4$Ni$_3$O$_{10}$ sample collected at $T = 5$~K [panel (a)] and $T = 110$~K [panel (b)].
(c)--(d) Fourier transforms of the data shown in panels (a) and (b). Solid lines represent the fit (red) and individual fit components. Dashed lines indicate the positions of internal field peaks expected from the 6-component fits.\\
(e) Temperature evolution of the Fourier transform spectra, showing the transition of the magnetic field distribution from a 5-peak to a 2-peak structure.
(f) Temperature dependence of the ZF-$\mu$SR signal fractions. The solid line corresponds to a fit of Eq.~\ref{eq:magnetic-fraction} to $f_{\rm m}(T)$.
(g) Temperature dependencies of the internal fields. The gray and pink lines in panels (f) and (g) indicate the positions of the magnetic ordering temperatures $T^\ast$ and $T_{\rm SDW}$, respectively.
}
 \label{fig:Ambient-pressure_muons}
\end{figure*}

\section{Ambient Pressure Studies of pristine ${\rm La}_4{\rm Ni}_3{\rm O}_{10-\delta}$} \label{sec:ambient-pressure}
%This section discusses a series of experiments conducted under ambient pressure conditions, including zero-field muon spin rotation/relaxation, x-ray, neutron powder diffraction (NPD), resistivity, and specific heat measurements.

\subsection{Zero-field $\mu$SR}

The zero-field (ZF) $\mu$SR response of La$_{4}$Ni$_{3}$O$_{10}$ measured at temperatures $T= 5$ and 110~K is presented in Figs.~\ref{fig:Ambient-pressure_muons}~(a)--(d). Panels (a) and (b) display the ZF asymmetry spectra, which represent the time evolution of the muon-spin polarization, while panels (c) and (d) show the Fourier transform of the data, illustrating the distribution of internal fields. Evidently, the time spectra and magnetic field distributions differ significantly between high and low temperatures. The number of peaks in the field distribution changes from five at $T=5$~K [Fig.~\ref{fig:Ambient-pressure_muons}~(c)] to two at $T=110$~K [Fig.~\ref{fig:Ambient-pressure_muons}~(d)]. The evolution of the magnetic field distribution is better visualized in the waterfall graph [Fig.~\ref{fig:Ambient-pressure_muons}~(e)], which shows that the five-peak structure transitions to a two-peak one at temperatures around 90~K.

Analysis revealed that the ZF-$\mu$SR time-spectra cannot be fitted  using simple cosine-type oscillating functions, which are typically used to describe a commensurate magnetic order\cite{Khasanov_La327_arxiv_2024, Amato-Morenzoni_book_2024, Yaouanc_book_2011, Blundell_book_2022}.  Instead, the magnetic field distributions shown in Figs.~\ref{fig:Ambient-pressure_muons}~(c) and (d) suggest the presence of long tails extending from the peak positions to approximately zero field. This feature is characteristic of incommensurate magnetic order, which is typically described by an Overhauser type distribution within the field domain  and by a zeroth-order Bessel function ($J_0$) within time domain \cite{Overhauser_JPhysChemSolids_1960, Schenck_PRB_2001, Amato_PRB_2014, Khasanov_PRB_MnP_2016}.

The fit of the ZF-$\mu$SR data was performed using Eq.~\ref{eq:asymmetry}. The magnetic (m) and nonmagnetic (nm) components, with corresponding weights $f_{\rm m}$ and $1-f_{\rm m}$, were described using Eqs.~\ref{eq:incommensurate} and \ref{eq:GKT}, respectively. The magnetic term (Eq.~\ref{eq:incommensurate}) required the presence of 6 components (5 oscillating and 1 fast relaxing) for the low-temperature data ($T\lesssim 90$~K), and 3 components (2 oscillating and 1 fast relaxing) for the high-temperature data ($T\gtrsim 90$~K).

The temperature evolution of magnetic volume fractions ($f_{\rm m,1}$ to $f_{\rm m,6}$) is shown in Fig.~\ref{fig:Ambient-pressure_muons}~(f). The total magnetic fraction $f_{\rm m}$ represents the sum of individual $f_{{\rm m,}i}$ components ($f_{\rm m}=\sum f_{{\rm m,}i}$). The solid line corresponds to the fit by means of the phenomenological expression:\cite{Khasanov_PRL_2008}
\begin{equation}
f_{\rm m}(T) = f_{\rm m}(0) \left[1 + \exp \left(\frac{T - T_{\rm SDW}}{\Delta T_{\rm SDW}}\right) \right]^{-1},
\label{eq:magnetic-fraction}
\end{equation}
where $f_{\rm m}(0)$ and $\Delta T_{\rm SDW}$ represent the zero-temperature value of the magnetic fraction and the width of the SDW transition, respectively. The fit yields $f_{\rm m}(0) = 0.98(1)$, $T_{\rm SDW} = 131.8(4)$~K, and $\Delta T_{\rm SDW} = 3.8(3)$~K. Consequently, below the SDW transition, nearly 100\% of the sample response is attributed to the magnetic contribution. The sharpness of the magnetic transition [$\Delta T_{\rm SDW} = 3.8(3)$~K] suggests that magnetism in La$_4$Ni$_3$O$_{10}$ sets in homogeneously.

The temperature dependence of the internal magnetic fields ($B_{\rm int,1}$ to $B_{\rm int,5}$) is shown in Fig.~\ref{fig:Ambient-pressure_muons}~(g). Two transitions, corresponding to changes in the internal fields at $T^\ast \simeq 80-90$~K and $T_{\rm SDW} \simeq 132$~K, are clearly observed. The first transition, at $T \simeq T_{\rm SDW}$, is characterized by the abrupt appearance of internal fields around $B_{\rm int} \sim 100$~mT. This behavior indicates that the magnetic transition at $T_{\rm SDW} \simeq 132$~K is first-order-like and is likely driven by a different kind of ordering, such as a structural transition (as observed in CrAs~\cite{Boller_SSC_1971, Khasanov_SciRep_2015}) or a nematic transition (as seen in Fe-based superconductors~\cite{Fernandes_NatPhys_2014}).
In La$_4$Ni$_3$O$_{10}$, the spin density wave and charge density wave transitions occur simultaneously, as reported in Ref.~\onlinecite{Zhang_NatCom_2020}. This suggests that the CDW transition may be the primary one, with charge modulation triggering the onset of SDW order.
It is worth noting that the change in the ordered volume fraction around $T_{\rm SDW}$, as measured by $\mu$SR and indicative of a first-order transition, would not be captured by neutron diffraction, which averages over the entire sample volume. As a result, the magnetic transition may appear smooth and second-order-like in neutron measurements [see, {\it e.g.}, Fig.~1(e) in Ref.~\onlinecite{Zhang_NatCom_2020}].

The second magnetic transition at $T^\ast \simeq 80-90$~K is characterized by a change in the magnetic field distribution from a two-peak to a five-peak structure.
Note, that the magnetic transitions at $T_{\rm SDW}$ and $T^\ast$ can be identified using different sets of fitting parameters. $T_{\rm SDW}$ is best determined from the temperature evolution of the magnetic volume fraction $f_m$ [Fig.~\ref{fig:Ambient-pressure_muons}~(f)]. In contrast, $B_{{\rm int},i}(T)$ [Fig.~\ref{fig:Ambient-pressure_muons}~(g)] does not approach zero due to the first-order nature of the magnetic transition. Conversely, $T^\ast$ is not reflected in the volume fraction dependencies but is associated with the loss of three out of five internal field components for temperatures exceeding 90~K. Both magnetic transitions are represented by the pink and gray vertical lines in Figs.~\ref{fig:Ambient-pressure_muons}~(f) and (g). A more systematic studies of the transition at $T^\ast$ were performed for the isotope substituted samples which are presented in the Section~\ref{sec:OIE}.

\subsection{Muon stopping sites and dipolar field calculations}

\begin{figure}[htb]
\includegraphics[width=0.9\linewidth]{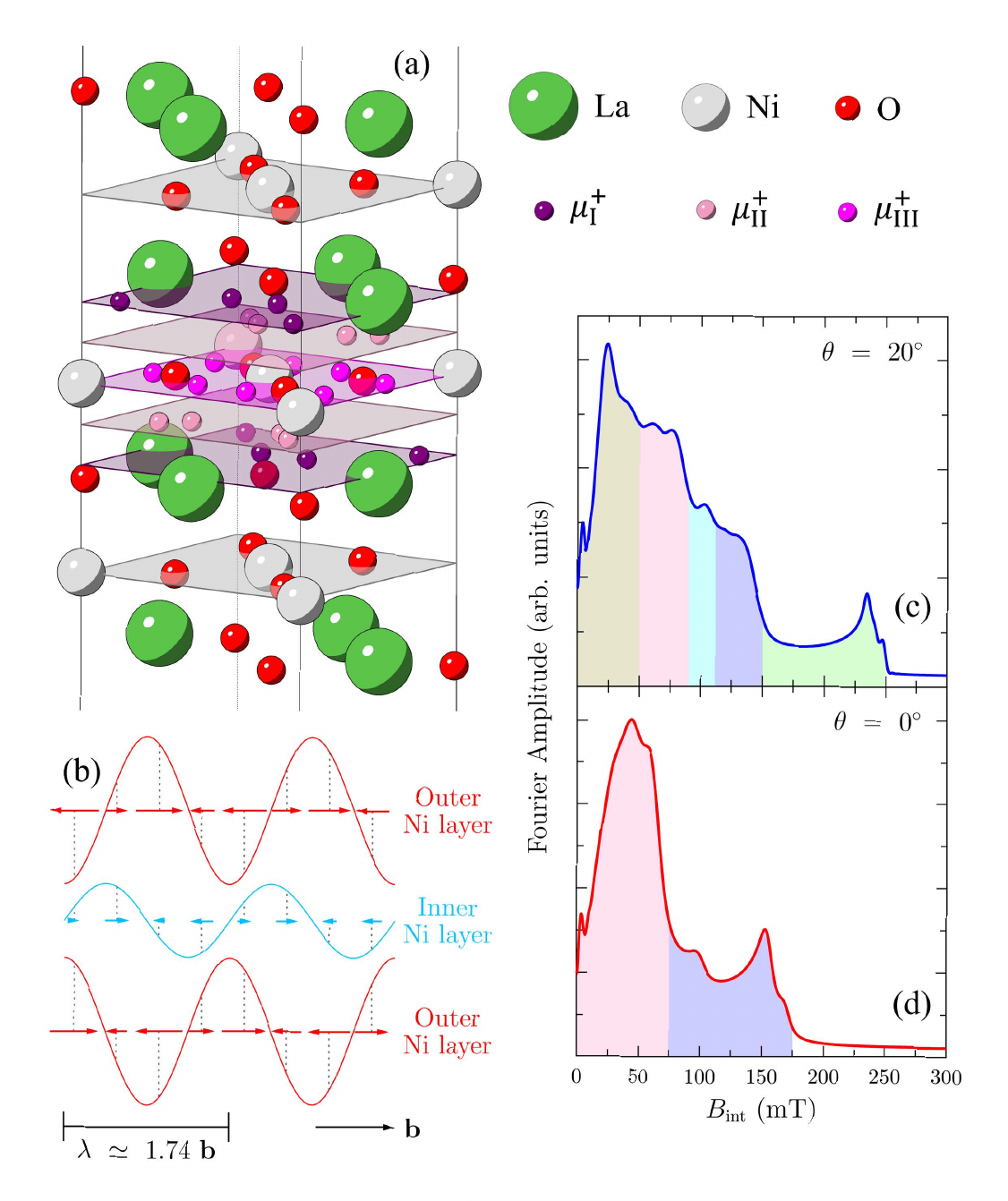}
\caption{%{\bf Muon stopping site calculations.}
(a) The three muon stopping sites with their relation to the Ni trilayer structure. Symmetrically equivalent positions are all shown, although in the experiment only one position is occupied at any one time.
(b) The magnetic structure used in simulations of the ZF $\mu$SR spectra. There is a large moment on the outer Ni layers, with a small one on the inner Ni layer. All moments point in the $\bm{b}$ direction, and form a SDW. There is a phase shift of 90$^\circ$ between each layer.
(c) and (d) Simulations of the dipole field at the muon stopping sites with the moments either pointing in the $a$b-plane ($\theta~=~0^\circ$), or rotated towards the $c$-axis ($\theta~=~20^\circ$). Colours under the curves correspond to the colours of the different fit components in Fig.~\ref{fig:Ambient-pressure_muons}~(c) and (d), highlighting the qualitative similarity.}
\label{fig:muon_sites}
\end{figure}

To understand the $\mu$SR data collected at ambient pressure, muon stopping site calculations were performed. Three distinct sites, listed in Table~\ref{table:muonSites}, were identified. Although these sites span a range of energies, all are likely to be occupied and show negligible displacement of the surrounding lattice, suggesting that the muon acts as a faithful probe of the system.
Each site is located in a different layer of the material, near the Ni trilayer structure, as illustrated in Fig.~\ref{fig:muon_sites}~(a). All three muon sites lie between 0.8 and 1~\AA\ from the nearest O atom, as is often found. The lowest-energy site is located in the La plane, the highest-energy site in the plane containing the inner Ni layer, and the intermediate-energy site lies in a plane between the other two.

\begin{table}
	\begin{tabular}{c|c|c}
		Muon site & Energy (eV) & Coordinates \\
		\hline
		I & 0 & (0.396, 0.904, 0.940) \\
		II & 0.18 & (0.204, 0.790, 0.530) \\
		III & 0.22 & (0.390, 0.696, 0.999) \\
	\end{tabular}
	\caption{The three identified muon stopping sites in La$_4$Ni$_3$O$_{10}$. The energies are given with respect to the lowest energy calculation. Fractional coordinates are given in terms of the conventional unit cell.}
	\label{table:muonSites}
\end{table}

\begin{figure*}[htb]
	\includegraphics[width=0.9\linewidth]{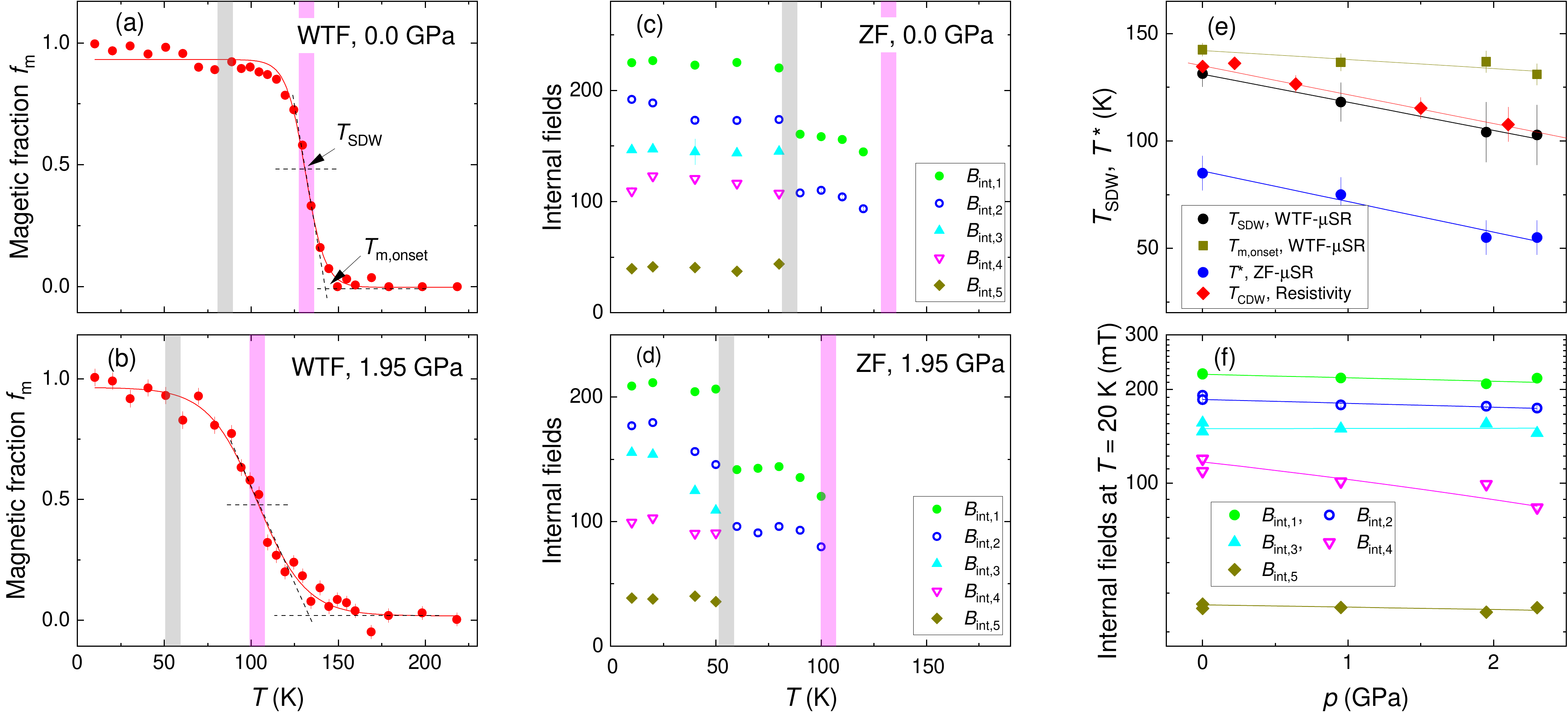}
\caption{
%{\bf Results of WTF and ZF-$\mu$SR under pressure experiments.}
(a)--(b) Temperature dependence of the magnetic volume fraction $f_{\rm m}$ measured in WTF-$\mu$SR experiments at $p = 0.0$~GPa [panel (a)] and $p = 1.95$~GPa [panel (b)]. The solid lines represent fits of Eq.~\ref{eq:magnetic-fraction} to the data.
(c)--(d) Temperature evolution of the internal fields measured in ZF-$\mu$SR experiments at $p = 0.0$~GPa [panel (c)] and $p = 1.95$~GPa [panel (d)]. The gray and pink lines in panels (a)--(d) indicate the positions of the magnetic ordering temperatures $T^\ast$ and $T_{\rm SDW}$, respectively.
(e) Pressure dependence of the magnetic ordering temperatures $T_{\rm m,onset}$, $T_{\rm SDW}$, and $T^\ast$, as well as the charge-density-wave ordering temperature $T_{\rm CDW}$. The solid lines are linear fits.
(f) Pressure dependence of the internal field components measured in ZF-$\mu$SR experiments at $T = 20$~K. The solid lines are linear fits.
}
	\label{fig:pressure_muons}
\end{figure*}

The dipole field at the muon stopping site were calculated by assuming all three muon sites are equally occupied. Note that very similar field distributions are produced by just considering the two lower-energy muon sites, therefore one cannot be certain whether or not the third site is occupied.
It was further assumed that there is no contribution to the spectra from the contact hyperfine interaction that would arise from overlap between the electronic and muon spin wavefunctions; this is consistent with our work on La$_3$Ni$_2$O$_7$~\cite{Khasanov_La327_arxiv_2024}.
As the magnetic structure is not fully understood, we have taken the structure proposed in Ref.~\onlinecite{Zhang_NatCom_2020} as a starting point, with an antiferromagnetically-coupled spin density wave on the outer two Ni layers, using the $q$-vector $\textit{\textbf{q}}=(0, 0.574, 0)$ obtained in our NPD measurements, see the Supplemental Materail part.\cite{Supplementa_Information} Note that this value of the $q-$vector stays relatively close to $\textit{\textbf{q}}=(0, 0.62, 0)$ reported by Zhang {\it et al.} in Ref.~\onlinecite{Zhang_NatCom_2020}.
Whilst this reproduces many of the features seen in the experiment, we find that this structure gives too much weight to fields near zero; to resolve this, we introduce a small moment on the inner Ni layer.
As $\mu$SR is not a $q$-resolved probe, we cannot categorically determine the ordering in this layer, hence we make the simplest assumption that the inner layer is also a SDW with the same wavevector, with a phase offset of $90^\circ$ from both outer layers, as shown in Fig.~\ref{fig:muon_sites}~(b).
We use a maximum moment of 2.81~$\mu_\textrm{B}$ on the outer Ni layers, consistent with the full moment of Ni, and the inner moment scaled to $\sim8\%$ of this value, as predicted by DFT calculations~\cite{LaBollita2024Electronic}.

Considering first the situation where the moments point in the $b$-direction, as predicted in Ref.~\onlinecite{Zhang_NatCom_2020}, our simulations qualitatively resemble the experimental data above 90~K, as shown in Fig.~\ref{fig:muon_sites}~(d).
The majority of muons experience a low internal field, with the remainder contributing to a continuous distribution, peaking maximally around 150~mT.
This suggests that at high-$T$, the magnetic structure is that of coupled SDWs, with the magnetic moments pointing in the $ab$-plane, and a smaller moment on the inner Ni layer.
At low-$T$ the spectra exhibit additional peaks, suggesting the field at the muon site becomes more varied.
There are myriad ways to break this degeneracy, we have considered two key scenarios.
(i) Increasing moment on the inner Ni layer.
This leads to many more peaks in the spectra, but leaves very little weight at lower magnetic fields, and therefore seems inconsistent with the experimental data.
(ii) Canting the moments out of the $ab$-plane.
There are many ways that this could occur, so we have considered the simplest case of rotating all the moments in the same way.
We show in Fig.~\ref{fig:muon_sites}~(c) the result when all the moments are rotated $\theta~=~20^\circ$ towards the $c$-axis.
This produces a qualitatively good match with the experimental data [Figs.~\ref{fig:Ambient-pressure_muons}~(c) and (d)], with the key features all reproduced with approximately the correct weighting.
As for $\theta~=~0^\circ$, the absolute magnitude of the internal field is not correct at lower fields, suggesting that, whilst the main features of the magnetic structure are well explained, there are some subtle details that a $q$-resolved probe will be best placed to explore.
Our two key results are that (i) there is likely a small magnetic moment on the inner Ni layers, and (ii) that it is most likely that the moments lie in the $ab$-plane above 90~K, before subtly distorting, gaining a $c$-axis component at lower $T$.

\section{Experiments under  Hydrostatic Pressure} \label{sec:pressure-experiments}
%This section discusses the muon-spin rotation/relaxation and resistivity experiments conducted under pressures up to $\simeq 2.3$~GPa.

\subsection{$\mu$SR experiments}

In $\mu$SR experiments conducted under pressure, a significant fraction of muons (approximately 50\%) stop in the pressure cell walls, contributing to the $\mu$SR response as background. Fitting the $\mu$SR data under pressure using Eq.~\ref{eq:asymmetry} requires knowledge of the background contribution from the pressure cell, which was determined in a separate set of experiments~\cite{Khasanov_HPR_2016}.
The analysis of ZF-$\mu$SR under pressure data was performed using Eqs.~\ref{eq:asymmetry}, \ref{eq:incommensurate}, and \ref{eq:GKT}. The $\mu$SR experiments conducted in a weak transverse field (WTF, $B_{\rm WTF}=5$~mT), \textit{i.e.}, with a field applied perpendicular to the initial muon-spin polarization, were analyzed using Eq.~\ref{eq:asymmetry}, with the sample part described by Eq.~\ref{eq:WTF}.

The temperature dependence of the magnetic volume fraction, obtained from WTF experiments at $p=0.0$ $1.95$~GPa, are shown in Figs.~\ref{fig:pressure_muons}~(a) and (b). Solid lines represent fits of Eq.~\ref{eq:magnetic-fraction} to the $f_{\rm m}(T)$ data. The onset of the magnetic ordering ($T_{\rm m,onset}$) is determined
as the crossing point of the linearly extrapolated $f_{\rm m}(T)$ line in  the vicinity of the magnetic transition with $f_{\rm m}=0$ line.
Temperature dependencies of internal fields, derived from fits to ZF-$\mu$SR data, are shown in Figs.~\ref{fig:pressure_muons}~(c) and (d).
The magnetic ordering temperatures $T^\ast$ and $T_{\rm SDW}$ are indicated by gray and pink vertical lines.

The pressure dependencies of $T_{\rm SDW}$ and $T_{\rm m,onset}$, as determined from WTF-$\mu$SR data, and $T^\ast$, obtained from changes in the magnetic field distribution -- specifically, the transition from a five-peak to a two-peak structure in ZF-$\mu$SR data -- are presented in Fig.~\ref{fig:pressure_muons}(e).
The solid lines represent linear fits:
\begin{align*}
  T_{\rm m,onset}(p) &= 142.1(1.5) - p\cdot4.2(1.2)~\text{K/GPa}, \\
  T_{\rm SDW}(p) &= 131.1(8) - p\cdot13.1(7)~\text{K/GPa}, \\
  T^\ast(p) &= 86(3) - p\cdot14(2)~\text{K/GPa}.
\end{align*}
The nearly identical pressure derivatives of $\simeq -13$~K/GPa for $T_{\rm SDW}(p)$ and $T^\ast(p)$ indicate that these transition temperatures decrease at a similar rate with increasing pressure. In contrast, $T_{\rm m,onset}$ decreases approximately three times more slowly.

The pressure-enhanced difference between $T_{\rm m,onset}$ and $T_{\rm SDW}$ suggests that magnetic fluctuations -- acting as precursors to static magnetic order and detectable by $\mu$SR -- are effectively enhanced under pressure. Indeed, the separation between $T_{\rm m,onset}$ and $T_{\rm SDW}$ increases from approximately 10~K at $p = 0.0$~GPa to about 30~K at $p \simeq 2.3$~GPa.

The pressure dependence of internal fields ($B_{\rm int,1}$ to $B_{\rm int,5}$) measured at $T=20$~K is shown in Fig.~\ref{fig:pressure_muons}~(f). Typically, the internal field at the muon-stopping site is proportional to the value of the ordered magnetic moments, {\it i.e.}, to the magnetic moments of Ni ions in our case. However, in La$_4$Ni$_3$O$_{10}$, the situation appears more complex. Linear fits of $B_{{\rm int},i}(p)$'s reveal that three internal fields ($B_{\rm int,1}$, $B_{\rm int,2}$, and $B_{\rm int,5}$) decrease with pressure at a similar rate, ${\rm d}\ln B_{{\rm int},i}/{\rm d}p \simeq -0.018(5)$~GPa$^{-1}$. In contrast, $B_{\rm int,3}$ remains nearly constant [${\rm d}\ln B_{{\rm int},3}/{\rm d}p = 0.003(10)~$GPa$^{-1}$], while $B_{\rm int,4}$ decreases six times faster [${\rm d}\ln B_{{\rm int},4}/{\rm d}p = -0.128(10)$~GPa$^{-1}$]. This suggests that pressure not only reduces the value of ordered magnetic moments, which would result in a similar pressure evolution for all internal fields, but may also lead to slight rearrangements of magnetic moments, causing different pressure dependencies for $B_{\rm int,3}$ and $B_{\rm int,4}$. Further studies, particularly precise determinations of the magnetic structure at various pressures, are needed to explain this observation.

\subsection{Resistivity measurements}

\begin{figure}[htb]
\includegraphics[width=0.8\linewidth]{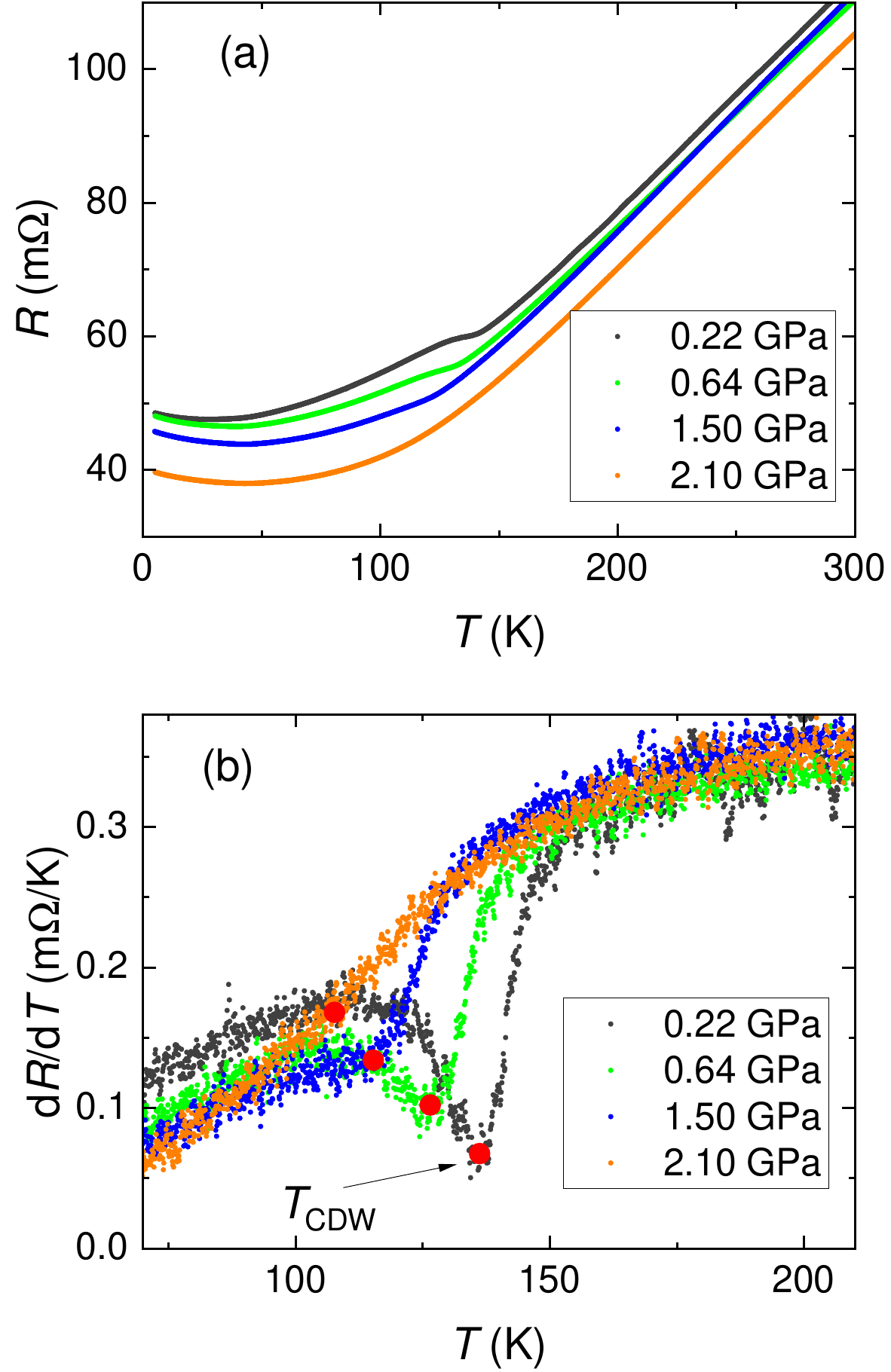}
\caption{%{\bf Pressure evolution of charge density wave (CDW) order.}
(a) Temperature dependencies of resistivity $R$ measured at pressures $p = 0.22$, 0.64, 1.50, and 2.10~GPa.
(b) First derivatives of the resistivity curves. Red dots indicate the CDW transition temperature $T_{\rm CDW}$.
%Error bars for $T_{\rm CDW}$ represent uncertainties in determining the transition temperature.
}
\label{fig:resistivity}
\end{figure}

The results of resistivity measurements at pressures ranging from 0.22 to 2.1~GPa are presented in Fig.~\ref{fig:resistivity}. The anomaly in the $R(T)$ data, which is associated in the literature with the appearance of the CDW ordered state, is visible at low pressures but nearly vanishes at $p \simeq 2.1$~GPa [Fig.~\ref{fig:resistivity}~(a)]. The data suggest that pressure suppresses both, the onset of CDW transition and the associated anomaly in $R(T)$ curves. To make CDW feature more apparent, Fig.~\ref{fig:resistivity}~(b) shows the first derivatives of the $R(T)$ data. The `dip' in ${\rm d}R(T)/{\rm d}T$ curves was associated with the CDW transition, and it is marked by red points in Fig.~\ref{fig:resistivity}~(b).

The pressure dependence of $T_{\rm CDW}$, combined with $T_{\rm m,onset}$, $T_{\rm SDW}(p)$ and $T^\ast(p)$, is shown in Fig.~\ref{fig:pressure_muons}~(e). A linear fit yields
\begin{align*}
T_{\rm CDW}(p) = 135.1(8)\,{\rm K} - p \cdot 13(2)~\text{K/GPa},
\end{align*}
in agreement with the $T_{\rm SDW}(p)$ data. This suggests that both transition temperatures are consistent within the experimental uncertainty and exhibit similar pressure dependencies.

\section{Oxygen-isotope substitution experiments} \label{sec:OIE}

\subsection{Oxygen-isotope effect on phonon modes}

The substitution of the lighter $^{16}{\rm O}$ isotope with the heavier $^{18}{\rm O}$ is expected to produce systematic downshifts of oxygen-dominated phonons. This effect is indeed observed, as shown in Figs.~\ref{fig:Characterization}~(e,f), where the Raman lines of $^{18}{\rm O}$-substituted La$_4$Ni$_3$O$_{10}$ appear at lower frequencies compared to those of the $^{16}{\rm O}$-substituted and pristine samples.

\begin{figure}[htb]
\includegraphics[width=0.9\linewidth]{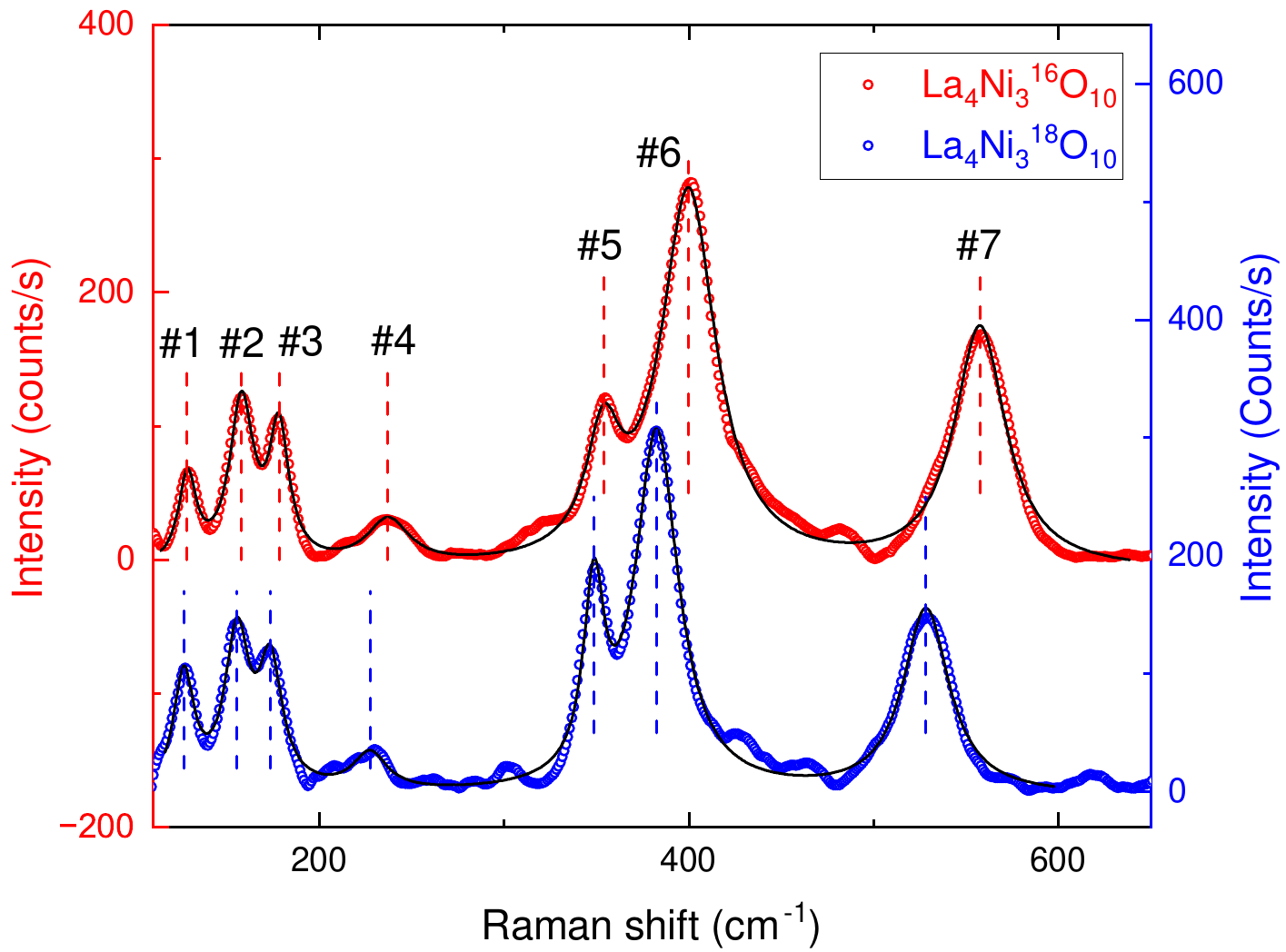}
\caption{
Room-temperature Raman spectra of oxygen-isotope substituted La$_4$Ni$_3$O$_{10}$ samples. Solid lines represent fits using seven Lorentzian functions. Dashed lines indicate the peak positions.}
 \label{fig:Raman_lines}
\end{figure}

The room-temperature Raman spectra of the $^{16}{\rm O}$- and $^{18}{\rm O}$-substituted La$_4$Ni$_3$O$_{10}$  samples were analysed by fitting the seven most pronounced Raman lines with Lorentzian functions, see Fig.~\ref{fig:Raman_lines}. Note that the spectra in Fig.~\ref{fig:Raman_lines} were collected at different laser-beam positions than those shown in Fig.~\ref{fig:Characterization}~(e). Variations in relative intensities arise from the polycrystalline nature of the sample. Because of the relatively small laser spot size ($\simeq 50$~$\mu$m), the Raman response is highly sensitive to the orientation of the illuminated grains. The Raman wavenumbers ($\nu^{16}$ and $\nu^{18}$), averaged over six different positions, are summarised in Table~\ref{table:Raman-results}.

\begin{figure}[htb]
\includegraphics[width=0.8\linewidth]{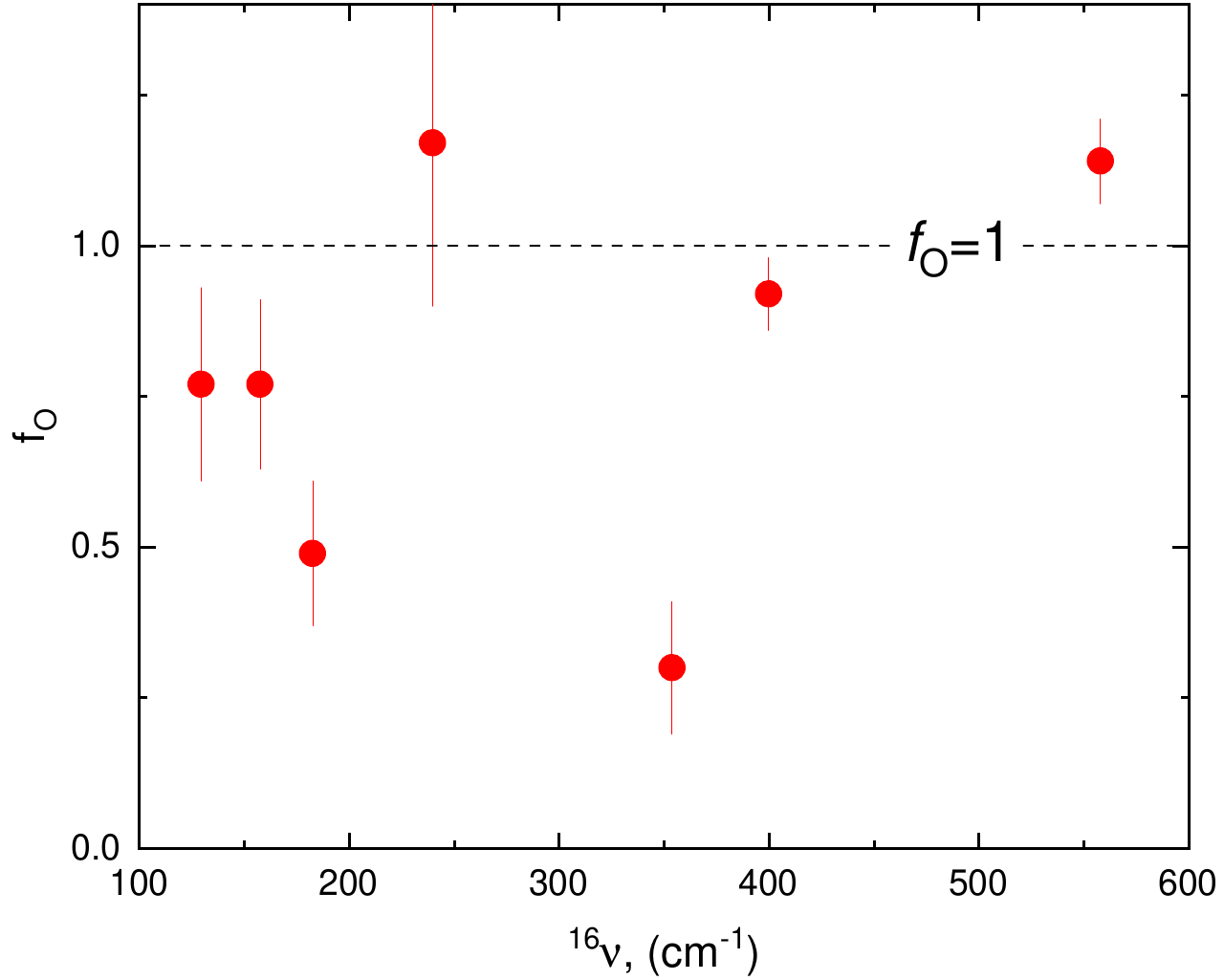}
\caption{
The oxygen participation parameter $f_{\rm O}$ as a function of the Raman wavenumber $\nu^{16}$.}
 \label{fig:f_O}
\end{figure}

The isotope dependence of the Raman wavenumbers $\nu^{16}$ and $\nu^{18}$ was analysed using a partial mass-scaling formalism developed in Refs.~\onlinecite{Cardona_RMP_2005, Menedez_Philmag_1994, Zhang_PRB_1997}:
\begin{equation}
\nu(x) \;=\; \nu^{16}\,\Bigg(\sqrt{\frac{16}{\bar M(x)}}\Bigg)^{f_{\rm O}},
\qquad
\bar M(x)=16(1-x)+18x.
\label{eq:nu_scaling}
\end{equation}
Here $x$ is the $^{18}$O fraction, $\nu^{16}$ is the frequency in the pure $^{16}$O sample, $\bar M$ is the average oxygen mass in the partially $^{18}$O-substituted sample, and $f_{\rm O}$ is the oxygen participation parameter. Taking logarithms gives
\begin{equation}
f_{\rm O} \;=\; \frac{\ln\!\big(\nu(x)/\nu^{16}\big)}{\ln\!\big(\sqrt{16/\bar M(x)}\big)}
\;=\;
\frac{\ln(1+s)}{\ln\!\big(\sqrt{16/\bar M(x)}\big)}~,
\label{eq:fO_def}
\end{equation}
with the fractional shift $s=\nu^{18}/\nu^{16}-1$. For an $^{18}$O isotope fraction of $x\simeq82$\%, the values of the oxygen participation parameter $f_{\rm O}$ were calculated and are summarised in Table~\ref{table:Raman-results} and Fig.~\ref{fig:f_O}.

\begin{table*}[t]
\centering
\caption{Raman-active phonon frequencies ($\nu^{16}$/$\nu^{18}$), fractional shift ($s$), and oxygen participation factor ($f_{\rm O}$) in $^{16}$O- and $^{18}$O-substituted La$_4$Ni$_3$O$_{10}$ at 300~K. Theoretical frequencies ($\nu^{\rm calc}$) and assigned phonon modes are from Ref.~\onlinecite{Suthar_arxiv_2025} (see Figs.~2f--j and the Supplemental Material therein).}
\begin{tabular}{c|c|c|c|c|c}
$\nu^{16}$ & $\nu^{18}$ &$s$&$f_O$&$\nu^{\rm calc}$& Assignment \\
(cm$^{-1}$)& cm$^{-1}$) &   &     &(cm$^{-1}$)      & \\
\hline
\multirow{2}{*}{129.3(1)}&\multirow{2}{*}{126.1(1)}&\multirow{2}{*}{-0.017(3)}&\multirow{2}{*}{0.76(15)} & \multirow{2}{*}{127.2\footnotemark[1]} & Collective bending involving apical\\
&&&&& and planar oxygens together with La/Ni.\footnotemark[1] \\
\multirow{2}{*}{157.4(1)}&\multirow{2}{*}{154.8(1)}&\multirow{2}{*}{-0.017(3)}&\multirow{2}{*}{0.75(12)} & \multirow{2}{*}{164.1\footnotemark[1]} & Ni and apical oxygen motion within \\
&&&&& Ni-O and $ab-$ planes.\footnotemark[1] \\
\multirow{2}{*}{182.5(1)}&\multirow{2}{*}{173.5(1)}&\multirow{2}{*}{-0.025(6)}&\multirow{2}{*}{0.48(11)} & \multirow{2}{*}{182.5\footnotemark[1]} & Mixed apical-oxygen displacements (in-plane and $c$),\\
&&&&& accompanied by Ni and La motion.\footnotemark[1] \\
\multirow{2}{*}{239.6(1)}&\multirow{2}{*}{223.9(1)}&\multirow{2}{*}{-0.054(12)}&\multirow{2}{*}{1.15(26)} & \multirow{2}{*}{239.6\footnotemark[1]} & Apical-oxygen dominated vibration along $c$, \\
&&&&& with additional planar-oxygen contribution.\footnotemark[1] \\
\multirow{2}{*}{353.7(1)}&\multirow{2}{*}{348.7(1)}&\multirow{2}{*}{-0.014(5)}&\multirow{2}{*}{0.29(10)} & \multirow{2}{*}{349.3\footnotemark[1]} & Apical-oxygen and Ni displacements (outer layers),\\
&&&&& mixed in-plane/$c$-axis character.\footnotemark[1] \\
\multirow{2}{*}{399.7(1)}&\multirow{2}{*}{382.7(1)}&\multirow{2}{*}{-0.043(3)}&\multirow{2}{*}{0.90(6)}  & \multirow{2}{*}{396.3\footnotemark[1]} & Planar-oxygen quadrupolar stretch (outer planes),\\
&&&&& vibrations within the $ab$ plane.\footnotemark[1] \\
\multirow{2}{*}{557.8(1)}&\multirow{2}{*}{528.4(1)}&\multirow{2}{*}{-0.053(3)}&\multirow{2}{*}{1.11(6)}  & \multirow{2}{*}{554.9\footnotemark[1]} & High-energy O stretch, dominated by apical/planar \\
&&&&& oxygen displacements mainly along $c-$axis.\footnotemark[1] \\
\end{tabular}

\footnotetext[1]{From DFPT calculations reported in Ref.~\onlinecite{Suthar_arxiv_2025} (see Figs.~2f--j and the Supplemental Information therein, where all Raman-active modes of La$_4$Ni$_3$O$_{10}$ are listed).}
\label{table:Raman-results}
\end{table*}

The experimentally observed Raman frequencies were assigned to the calculated phonon modes reported in Ref.~\onlinecite{Suthar_arxiv_2025} and its Supplementary Information (see the last two columns of Table~\ref{table:Raman-results}). Comparison of the measured and theoretical values leads to the following conclusions: \\
(i) The Raman lines near 558, 400, and 240~cm$^{-1}$ are assigned predominantly to oxygen vibrations, which is consistent with the oxygen participation factor being close to unity ($f_{\rm O}\simeq 1.0$, see Fig.~\ref{fig:f_O}). \\
(ii) Modes involving La and Ni motion (near 129, 157, 183, and 354~cm$^{-1}$) exhibit reduced oxygen participation, with $f_{\rm O}<1$.

It is noteworthy that for the oxygen-dominated modes (near 558, 400, and 240~cm$^{-1}$), both apical and planar oxygens contribute to the vibrations, resulting in $f_{\rm O}\simeq 1.0$ (see the last two columns in Table~\ref{table:Raman-results} and Fig.~\ref{fig:f_O}). This strongly suggests a uniform distribution of $^{18}$O throughout the sample. The ratio of $\simeq 20$\% $^{16}$O to $\simeq 80$\% $^{18}$O appears to remain constant across all oxygen sites (apical, inner planes, and outer planes).

\subsection{Oxygen-isotope effect on the Spin-Density Wave transitions} \label{sec:OIE_muSR}

\begin{figure*}[htb]
\includegraphics[width=1.0\linewidth]{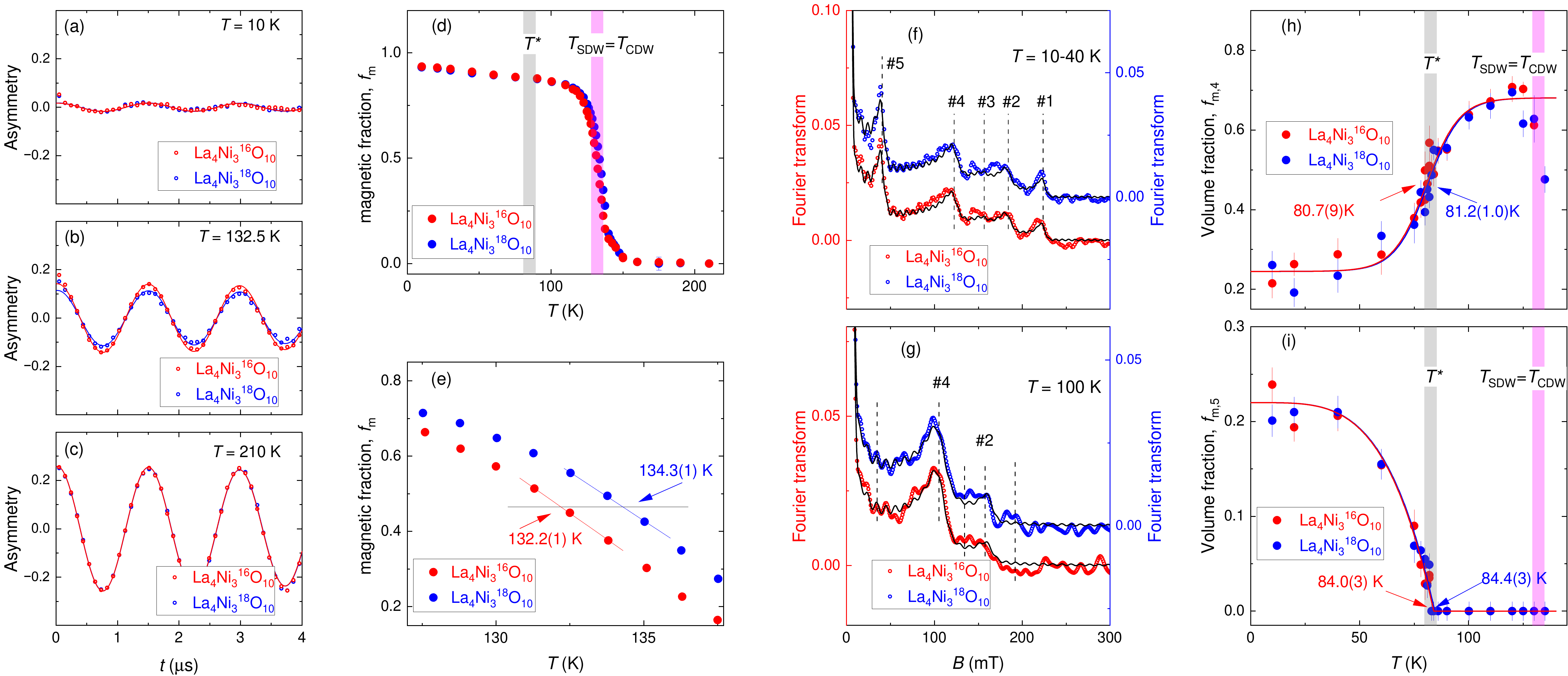}
\caption{
%{\bf Oxygen isotope effect on SDW transition temperatures.}
(a)--(c) WTF-$\mu$SR time spectra of $^{16}$O/$^{18}$O-substituted La$_4$Ni$_3$O$_{10}$ samples measured at $T = 10$~K, 132~K, and 210~K.
(d) Temperature dependence of the magnetic volume fraction $f_{\rm m}$ for $^{16}$O/$^{18}$O-substituted La$_4$Ni$_3$O$_{10}$ samples. The grey and pink stripes represent $T_{\rm SDW} = T_{\rm CDW}$ and $T^\ast$, respectively, as obtained from $\mu$SR and resistivity studies of the pristine La$_4$Ni$_3$O$_{10}$ sample [see Figs.~\ref{fig:Ambient-pressure_muons} and \ref{fig:pressure_muons}, and Sections~\ref{sec:ambient-pressure} and \ref{sec:pressure-experiments}].
(e) Expanded view of the $f_{\rm m}(T)$ curves in the vicinity of the SDW transitions.
(f)--(g) Distribution of internal fields in $^{16}$O/$^{18}$O-substituted La$_4$Ni$_3$O$_{10}$ samples, collected at temperatures ranging from 10 to 40~K [panel (f)] and at $T = 100$~K [panel (g)]. Dashed lines represent the positions of the five field components.
(h) Temperature dependence of the magnetic volume fraction $f_{\rm m,4}$. Solid lines are fits to the data using Eq.~\ref{eq:frac4}.
(i) Temperature dependence of $f_{\rm m,5}$. Solid lines represent fits using Eq.~\ref{eq:frac5}.
}
 \label{fig:OIE_muSR}
\end{figure*}

Figures~\ref{fig:OIE_muSR}(a)--(c) present time spectra collected in WTF-$\mu$SR experiments for $^{16}$O- and $^{18}$O-substituted La$_4$Ni$_3$O$_{10}$ samples. At $T = 10$~K [panel (a)], the oscillations in $B_{\rm WTF}$ are strongly suppressed, indicating that both samples remain in the magnetic state. At $T = 132.5$~K [panel (b)], the asymmetries are partially recovered, reaching approximately half of the full amplitude, with a slightly higher absolute value for the La$_4$Ni$_3\,^{16}$O$_{10}$ sample compared to the La$_4$Ni$_3\,^{18}$O$_{10}$ one. This suggests that the magnetic volume fraction $f_{\rm m}$ is higher in the $^{18}$O-substituted sample than in the $^{16}$O-substituted one.
Above the SDW transition [$T = 210$~K, panel (c)], the asymmetries for both samples become equal and reach a maximum value of approximately 0.27, consistent with the characteristics of the GPS $\mu$SR spectrometer.\cite{Amato_RSI_2017}

The temperature evolution of the magnetic volume fractions obtained from fits to the WTF-$\mu$SR data is presented in Fig.~\ref{fig:OIE_muSR}(d). Figure~\ref{fig:OIE_muSR}(e) shows an expanded view of the $f_{\rm m}(T)$ curves in the vicinity of $T_{\rm SDW}$. The $T_{\rm SDW}$ transition temperatures were estimated from the intersection of the linearly extrapolated $f_{\rm m}(T)$ curves near $T_{\rm SDW}$ with the reference line $f_{\rm m}(T) = 0.5 \cdot f_{\rm m}(10~{\rm K})$. The fits yield an isotope shift of $T_{\rm SDW}$ as $T_{\rm SDW}^{18} - T_{\rm SDW}^{16} = 2.1(1)$~K.

The OIE on the spin-reorientation transition temperature $T^\ast$ was further investigated in ZF-$\mu$SR experiments. The magnetic field distributions measured in the pristine La$_4$Ni$_3$O$_{10}$ sample, shown in Figs.~\ref{fig:Ambient-pressure_muons}~(c)–(e), as well as those for the isotope-substituted samples in Figs.~\ref{fig:OIE_muSR}~(f) and (g), suggest that the spin-reorientation transition at $T^\ast$ is characterized by the disappearance of components \#1, \#3, and \#5. The most pronounced changes occur in component \#5, which is among the strongest below $T^\ast$ but disappears above it, and in component \#4, whose intensity nearly doubles above $T^\ast$. The corresponding temperature dependencies of the volume fractions $f_{\rm m,4}$ and $f_{\rm m,5}$ are shown in Figs.~\ref{fig:OIE_muSR}~(h) and (i).

The solid lines in Figs.~\ref{fig:OIE_muSR}~(h) and (i) correspond to fits of the following models to the $f_{\rm m,4}(T)$ and $f_{\rm m,5}(T)$ data:
\begin{equation}
f_{\rm m,4}(T) = f_{\rm m,0} - \Delta f_{\rm m,4} \left[1 + \exp \left(\frac{T - T^\ast}{\Delta T^\ast}\right) \right]^{-1},
\label{eq:frac4}
\end{equation}
and
\begin{equation}
f_{\rm m,5}(T) =
\begin{cases}
f_{\rm m,high} \; \; , & T > T^\ast \\[5pt]
f_{\rm m,high}+\Delta f_{\rm m,5} \left[1-\left(\frac{T}{T^\ast}\right)^n\right] , & T < T^\ast
\end{cases}.
\label{eq:frac5}
\end{equation}
Here $f_{\rm m,0}$ and $f_{\rm m,high}$ represent the volume fractions at $T=0$ and $T>T^\ast$, respectively, and $\Delta f_{\rm m} = f_{\rm m,0} - f_{\rm m,high}$ denotes the transition amplitude. $\Delta T^\ast$ characterizes the width of the transition, while $n$ is the exponent.
To reduce correlation between parameters, the fits were performed by assuming equal values of $f_{\rm m,0}$, $\Delta f_{\rm m}$, and $\Delta T^\ast$ in Eq.~\ref{eq:frac4}, and equal values of $f_{\rm m,high}$ and $\Delta f_{\rm m}$ in Eq.~\ref{eq:frac5} for both $^{16}{\rm O}$- and $^{18}{\rm O}$-substituted samples, while keeping $T^\ast$ isotope-dependent. The fits confirm the absence of a measurable isotope shift in the spin-reorientation transition temperature $T^\ast$ within the experimental accuracy. The estimated shifts in the $T^\ast$ transition temperatures were found to be $ T^{\ast,18} - T^{\ast,16} = 0.5(1.4)$~K and 0.4(5)~K from the fits to $f_{\rm m,4}(T)$ and $f_{\rm m,5}(T)$ data sets, respectively.

It should be noted that the determination of the spin-reorientation transition temperature $T^\ast$ depends on the choice of the model function -- Eqs.~\ref{eq:frac4} and \ref{eq:frac5} in our case -- which may introduce relatively high uncertainties in the absolute value of $T^\ast$ [$\simeq$81 or $\simeq$84~K, as shown in Figs.~\ref{fig:OIE_muSR}~(h) and (i)].
On the other hand, fitting Eqs.~\ref{eq:frac4} and \ref{eq:frac5} to $f_{\rm m,4}(T)$ and $f_{\rm m,5}(T)$ data yields considerably smaller statistical errors, implying that uncertainties in the relative change of $T^\ast$ are low.

\begin{table*}[htb]
\caption{The spin density wave ($T_{\rm SDW}$ and $T^\ast$) and the charge density wave ($T_{\rm CDW}$) transition temperatures obtained at pristine and oxygen-isotope-substituted La$_4$Ni$_3$O$_{10}$ samples. $T_{tr}^{18}-T_{tr}^{16}$ denotes isotope shift of the transition temperature  ($T_{tr}= T_{\rm CDW}$, $T^\ast$, or $T_{\rm SDW}$). The oxygen isotope exponent $\alpha=-{\rm d}\ln T_{tr}/{\rm d}\ln M$ ($M$ is the mass of the oxygen isotope) is corrected for uncomplete isotope exchange $\simeq 82$\%.  }
\begin{tabular}{c|c|c|c|c|c|c }
%  \hline
% \cline{2-11}
    &La$_4$Ni$_3$O$_{10}$ &La$_4$Ni$_3\, ^{16}$O$_{10}$ &La$_4$Ni$_3\, ^{18}$O$_{10}$& $T_{tr}^{18}-T_{tr}^{16}$ &$\alpha$&Technique  \\
    &(K)           & (K)          &(K)& (K)       &         &          \\
  \hline
$T_{\rm CDW}$&136.0(1)&135.8(1)&138.0(1)&2.2(1)&-0.16(1)&Resistivity\\
&&&&&&\\
%\hline
\multirow{2}{*}{$T_{\rm SDW}$}&131.3(4)&132.2(1)&134.3(1)&2.1(1)&-0.15(1)&WTF-$\mu$SR\\
              &131.8(4)&&&&&ZF-$\mu$SR\\
&&&&&&\\
%\hline
\multirow{2}{*}{$T^\ast$}&82.2(2.4)\footnotemark[1] &80.7(9)\footnotemark[1] &81.2(1.0)\footnotemark[1] &0.5(1.3)&-0.06(16)&ZF-$\mu$SR\\
              &83.1(1.3)\footnotemark[2] &84.0(3)\footnotemark[3] &84.4(3)\footnotemark[3]  &0.4(0.4)&-0.05(5) &ZF-$\mu$SR\\
%\hline
\end{tabular}
\footnotetext[1]{From the fit of Eq.~\ref{eq:frac4} to $f_{\rm m,4}(T)$ data.}
\footnotetext[3]{From the fit of Eq.~\ref{eq:frac5} to $f_{\rm m,1}(T)$ data.}
\footnotetext[3]{From the fit of Eq.~\ref{eq:frac5} to $f_{\rm m,5}(T)$ data.}
\label{table:OIE-results} % is used to refer this table in the text
\end{table*}

The magnetic ordering temperatures determined for the pristine and $^{16}$O/$^{18}$O-substituted samples investigated in the present study are summarized in Table~\ref{table:OIE-results}.
The temperature dependencies of $f_{\rm m}(T)$, and individual volume fraction components [$f_{\rm m,1}(T)$ to $f_{\rm m,5}(T)$]for pristine La$_4$Ni$_3$O$_{10}$, obtained from WTF- and ZF-$\mu$SR studies, are presented in the Supplemental Material \cite{Supplementa_Information}. The larger uncertainties in $T_{\rm SDW}$ and $T^\ast$ for the pristine sample arise from the coarser temperature steps used in the corresponding $\mu$SR experiments. Despite these larger uncertainties, the value of $T_{\rm SDW}$ for the pristine sample remains closer to that of the $^{16}$O-substituted sample than to the $^{18}$O-substituted one, as might be expected. The spin-reorientation temperature $T^\ast$ for the pristine sample also exhibits a higher uncertainty compared to the values obtained from fine temperature-step measurements on La$_4$Ni$_3\,^{16}$O$_{10}$ and La$_4$Ni$_3\,^{18}$O$_{10}$, but it remains within the same temperature range.

\subsection{Oxygen-isotope effect on the Charge-Density Wave transitions}

\begin{figure}[htb]
\includegraphics[width=0.8\linewidth]{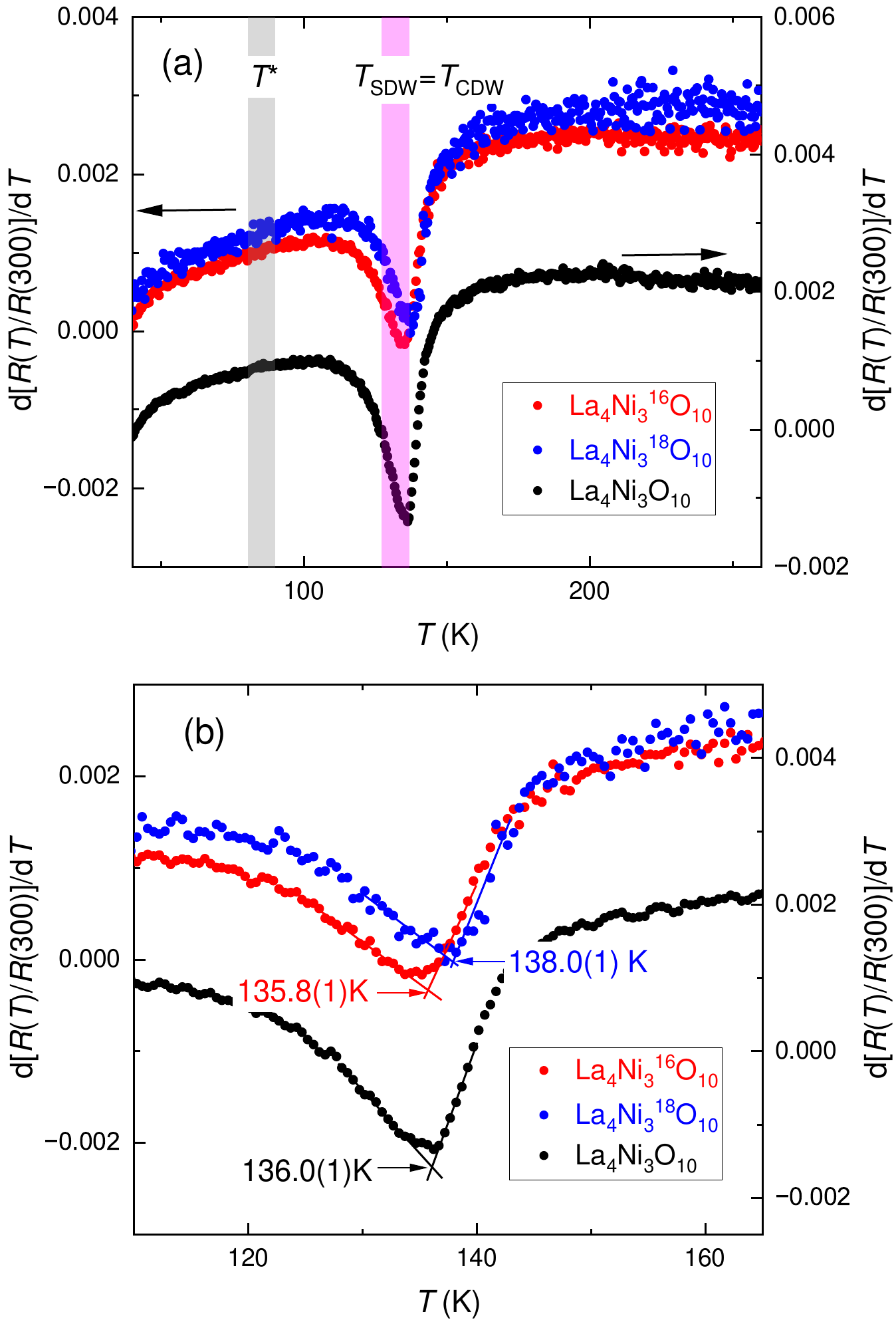}
\caption{
%{\bf Oxygen isotope effect on SDW transition temperatures.}
(a) First derivatives of the resistivity curves for pristine and isotope-substituted La$_4$Ni$_3$O$_{10}$ samples.\\
(b) Expanded view of the derivative curves in the vicinity of the CDW transition.
}
 \label{fig:OIE_resistivity}
\end{figure}

The oxygen isotope effect (OIE) on the CDW transition temperature was investigated by resistivity measurements. Figure~\ref{fig:Characterization}~(f) shows the temperature dependence of the resistivity normalized to its 300~K value, $R(T)/R(300)$. The raw resistivity curves exhibit weakly pronounced features at $T \sim 130$~K for all three La$_4$Ni$_3$O$_{10}$ samples. A more distinct structure becomes evident upon taking the first derivative of the resistivity data, as shown in Fig.~\ref{fig:OIE_resistivity}~(a).

Figure~\ref{fig:OIE_resistivity}(b) presents an expanded view of the resistivity derivative curves near the CDW ordering temperatures. The CDW transition temperature was determined from the intersection point of linear fits to the derivative curves near the local minimum. Clearly, the CDW transition temperatures for the pristine and $^{16}$O-substituted La$_4$Ni$_3$O$_{10}$ samples remain nearly the same, while $T_{\rm CDW}$ for La$_4$Ni$_3\,^{18}$O$_{10}$ is shifted to higher temperatures. The $T_{\rm CDW}$ values for the pristine and oxygen-isotope-substituted samples are summarized in Table~\ref{table:OIE-results}.
%The isotope exponent on the transition temperature $T_{tr}$ ($T_{\rm SDW}$, $T^\ast$, and $T_{\rm SDW}$), is defined as
%\begin{equation}
%\alpha_{tr} = -\frac{{\rm d}\ln T_{tr}}{{\rm d} \ln M},
%  \label{eq:isotope-exponent}
%\end{equation}
%and it is corrected for uncomplete isotope exchange $\simeq 82$\%.

\section{Conclusions and Outlook} \label{sec:Conclusions}

This study investigates the three-layer Ruddlesden–Popper nickelate La$_4$Ni$_3$O$_{10}$ using $\mu$SR and resistivity, offering comprehensive insights into the interplay between spin and charge orders under both ambient and high-pressure conditions, as well as the effects induced by substituting $^{16}$O with its heavier isotope $^{18}$O.

At ambient pressure, two magnetic transitions are identified at $T_{\rm SDW} \simeq 132$~K and $T^\ast \simeq 80$--$90$~K. The incommensurate nature of both magnetic states is confirmed by zero-field $\mu$SR experiments through the observation of an Overhauser-type magnetic field distribution in the frequency domain (corresponding to a zeroth-order Bessel function in the time domain). An abrupt onset of internal fields indicates a first-order-like character of the transition at $T_{\rm SDW}$.

Muon stopping site and dipole-field calculations suggest the presence of a small magnetic moment on the inner Ni layers and indicate that the magnetic moments likely lie within the $ab$-plane above $T^\ast$, gradually acquiring a $c$-axis component at lower temperatures.

The CDW order is evidenced by an anomaly in the resistivity curve, with $T_{\rm CDW}$ coinciding with $T_{\rm SDW}$, consistent with previous studies~\cite{Zhang_NatCom_2020}.

Under applied pressure, both resistivity and $\mu$SR measurements reveal the suppression of all transitions, including the two magnetic transitions ($T_{\rm SDW}$ and $T^\ast$) and the CDW transition. The pressure coefficients for these transitions are nearly identical, $\simeq -13$~K/GPa, and the SDW and CDW transitions remain closely intertwined under increasing pressure. The onset of magnetic ordering ($T_{\rm m,onset}$) decreases more slowly than the SDW ordering temperature, suggesting that magnetic fluctuations -- acting as precursors to static magnetic order and detectable by $\mu$SR -- are effectively enhanced under pressure.

\begin{figure}[htb]
\includegraphics[width=0.95\linewidth]{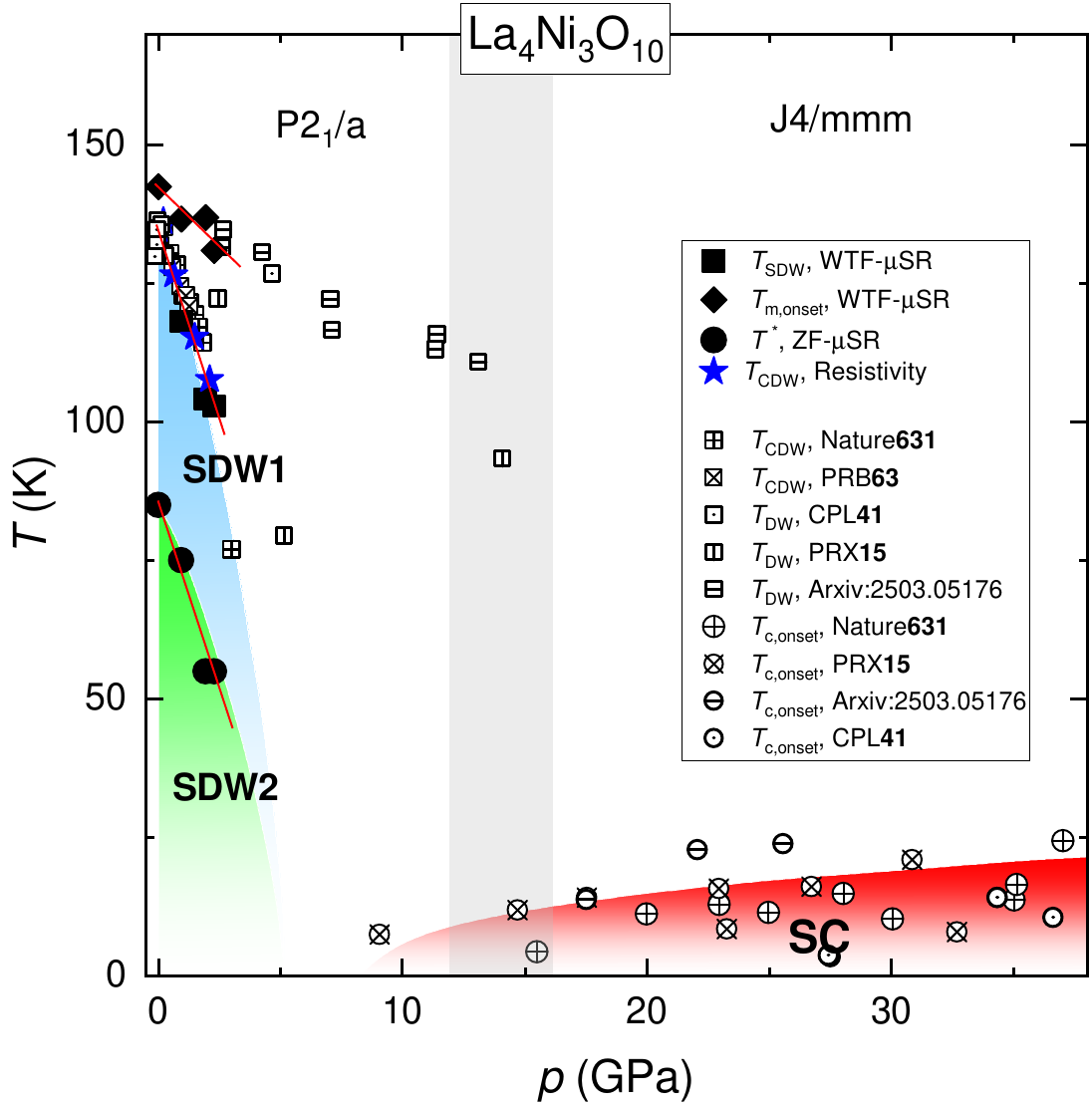}
\caption{The pressure-temperature ($p-T$) phase diagram of La$_4$Ni$_3$O$_{10}$.
Solid symbols represent the pressure dependencies of the magnetic ordering temperatures $T_{\rm m,onset}$, $T_{\rm SDW}$ and $T^\ast$, as obtained from ZF and WTF-$\mu$SR experiments (black squares and circles) and the CDW ordering temperature ($T_{\rm CDW}$) derived from resistivity studies (blue stars) in the present work. The open squares and circles denote the $T_{\rm CDW}$ {\it vs.} $p$ and the onset of the superconducting transition temperature $T_{\rm c}$ {\it vs.} $p$ data from Refs.~\onlinecite{Zhu_Nature_2024, Wu_PRB_2001, Zhang_arxiv_2023, Li_CPL_2024, Xu_arxiv_2025}.
}
 \label{fig:Phase-Diagram}
\end{figure}

The resulting phase diagram, presented in Fig.~\ref{fig:Phase-Diagram}, highlights the suppression of density-wave transitions with increasing pressure. Interestingly, $T_{\rm m,onset}$ appears to follow the pressure dependence of the density-wave transition observed in ultrafast optical pump-probe spectroscopy experiments.\cite{Xu_arxiv_2025}
Considering the difference in time scales between $\mu$SR and femtosecond optical spectroscopy ($10^{-5}$–$10^{-10}$\,s for $\mu$SR vs.\ $\sim 10^{-15}$\,s for femtosecond spectroscopy), one may assume that magnetic fluctuations which remain dynamic on the muon time scale could be detected as static order at the femtosecond scale. We cannot exclude, however, that the coincidence between $T_{\mathrm{m,onset}}$ and $T_{\mathrm{DW}}$ reported in Ref.~\onlinecite{Xu_arxiv_2025} is accidental.

The simultaneous suppression of density-wave transitions in La$_4$Ni$_3$O$_{10}$ contrasts with the behavior of the two-layer RP nickelate La$_3$Ni$_2$O$_7$~\cite{Khasanov_La327_arxiv_2024}, where external pressure enhances the separation between the SDW and CDW transitions. The differing pressure responses of La$_3$Ni$_2$O$_7$ and La$_4$Ni$_3$O$_{10}$ arise from the fact, that in La$_3$Ni$_2$O$_7$ the SDW and CDW orders are decoupled, allowing the SDW transition to be independently enhanced under pressure. Lattice compression likely strengthens exchange interactions between neighboring Ni moments, thereby increasing the SDW transition temperature.

In contrast, in La$_4$Ni$_3$O$_{10}$ the SDW and CDW orders are strongly coupled and intertwined~\cite{Zhang_NatCom_2020}, meaning that the SDW transition is highly dependent on the stability of the CDW. As pressure suppresses the CDW order, it simultaneously destabilizes the SDW, leading to a reduction in the SDW transition temperature. This contrast highlights the critical role of SDW–CDW coupling in determining the pressure response: La$_4$Ni$_3$O$_{10}$ exhibits mutual destabilization of both orders under pressure, whereas La$_3$Ni$_2$O$_7$ allows the SDW order to evolve more independently. It should be noted, however that ultrafast optical studies on both La$_3$Ni$_2$O$_7$, Ref.~\onlinecite{Meng_NatCom_2024} and La$_4$Ni$_3$O$_{10}$, Ref.~\onlinecite{Xu_arxiv_2025}, show that SDW order is suppressed
prior to the onset of superconductivity. This suggests that, while the double- and triple-layer nickelates display distinct behaviors at low pressures, their phase diagrams may converge as one approaches the superconducting regime at higher pressures.

The interpretation on the critical role of SDW–CDW coupling is further supported by oxygen isotope effect (OIE) experiments. In La$_4$Ni$_3$O$_{10}$, where the SDW and CDW orders are coupled, the OIE on both $T_{\rm SDW}$ and $T_{\rm CDW}$ is nearly identical. However, when the magnetic transition is decoupled from the CDW order -- as is the case for the spin-reorientation transition at $T^\ast$ -- the isotope effect on the corresponding magnetic ordering temperature remains negligible.
In La$_4$Ni$_3$O$_{10}$, where SDW and CDW orders are coupled and both break lattice translational symmetry, the OIE on both $T_{\rm SDW}$ and $T_{\rm CDW}$ is nearly identical. However, when the magnetic transition is decoupled from the CDW order -- as in the case of the spin-reorientation transition at $T^\ast$ -- no additional lattice translational symmetry is broken, and the isotope effect on the corresponding magnetic ordering temperature remains negligible.

The relationship between spin and charge orders in La$_4$Ni$_3$O$_{10}$ mirrors the intricate interplay observed in cuprates~\cite{Tranquada_Nature_1995, Ghiringhelli_Sciense_2012, Chang_NatPhys_2012, Fradkin_RMP_2015, Keimer_Nature_2015, Kivelson_RMP_2003, Guguchia_PRL_2020} and hole-doped nickelates such as La$_{2-x}$Sr$_x$NiO$_4$~\cite{Zhang_PNAS_2016, Zhang_PRL_2019}. In these systems, spin and charge orders are strongly intertwined and exhibit similar responses to external perturbations. However, important open questions remain: Why does the magnetic order change from commensurate to incommensurate as one moves from the two-layer to the three-layer RP nickelate system? What drives the pressure-enhanced separation of SDW and CDW transitions observed in La$_3$Ni$_2$O$_7$, in contrast to the intertwined behavior seen in La$_4$Ni$_3$O$_{10}$? Addressing these questions will require further experimental and theoretical investigations into the electronic structure and magnetic interactions of RP nickelates under varying conditions.

While finalizing this paper, we became aware of a report on ambient-pressure $\mu$SR studies of La$_4$Ni$_3$O$_{10}$~\cite{Cao_Arxiv_2025}, which reports the observation of a second magnetic transition around 70--100~K, accompanied by a change in the magnetic field distribution.

\section*{ACKNOWLEDGEMENTS}
R.K. acknowledges the authors of Ref.~\onlinecite{Suthar_arxiv_2025} for providing theoretical Raman data prior to publication, M. Hepting for helpful discussions, and J. Mitchell and D. Phelan for supplying neutron data on Pr$_4$Ni$_3$O$_{10}$. Z.G. acknowledges support from the Swiss National Science Foundation (SNSF) through SNSF Starting Grant (No. TMSGI2${\_}$211750). I.P. acknowledges financial support from Paul Scherrer Institute research grant No. 2021\_0134. D.J.G. acknowledges support from the Swiss National Science Foundation (SNSF) through Grant No. 200021E\_238113.

\appendix

\setcounter{figure}{0}
\renewcommand{\figurename}{Fig.}
\renewcommand{\thefigure}{A\arabic{figure}}

\section{Materials and Methods}

\subsection{Sample Preparation}

The La$_4$Ni$_3$O$_{10}$ polycrystalline sample was synthesized using a combination of mechanosynthesis and solid-state reaction method, similar to that described in Ref.~\onlinecite{Khasanov_La327_arxiv_2024}. The starting materials were La$_2$O$_3$ (5N, Sigma-Aldrich) and NiO (4N8, Alfa Aesar). Prior to synthesis, La$_2$O$_3$ and NiO were dried at 1200$^\circ$C and 280$^\circ$C, respectively, for 12 hours. Stoichiometric amounts of freshly dried La$_2$O$_3$ (21.490~g, 65.96 mmol) and NiO (7.390~g, 10.0~mmol) were well mixed in RM 100 (Retsch$^\circledR$) mechanical mortar. The mixture of oxides was divided into two aliquots and placed together with 10 mm balls in a 45 ml agate grinding bowl and then subjected to mechanosynthesis in a PULVERISETTE 7 planetary micro ball-mill (Fritsch GmbH) during 20 cycles at 600 rpm for 5 min milling and 5 min cooling conditions. The resulting black precursor powder was cold-pressed into a pellet by applying 4000~bar uniaxial pressure and annealed in oxygen (5N, PabGas) under 200~ml/min flow at 1100$^\circ$C for 10 h, followed by a final thermal treatment  performed at 500$^\circ$ for 6~h.

Approximately  1.0~g of La$_4$Ni$_3$O$_{10}$ initial material in a powder from was used for the isotope exchange process. The sample was divided into two equal parts ($\simeq0.5$~g each) and placed in ampules filled with $^{18}$O$_2$ (Euriso-Top, 97.1\% enrichment) and a mixture of natural isotopes of oxygen, hereafter referred to $^{16}{\rm O}_2$ (PanGas, 5N), respectively. The experimental setup is described in Refs.~\onlinecite{Conder_MatScIng_2001, Conder_PhysicaC_2023}.
The samples were annealed simultaneously in $^{16}$O$_2$ and $^{18}$O$_2$ atmospheres in small overpressure (up to $\sim1.5$~bar) at $T = 1000{^\circ}$C for 6 hours, followed by a post-annealing step at $T = 500{^\circ}$C for 24 hours.
To increase the isotope content in the $^{18}{\rm O}$-substituted samples, $^{18}{\rm O}_2$ gas was replaced and the IOE procedure was repeated four times. Since the mass spectrometer (MS) is coupled to the isotopic oxygen reactor, the isotope exchange rate can be monitored in situ and estimated based on simple reaction progress and equilibrium constants dependencies.\cite{Atkins_Physical_Cehmistry} The $^{18}{\rm O}$ content in La$_4$Ni$_3 ^{18}$O$_{10}$ sample is thus derived to be 82(2)\%.

\subsection{$\mu$SR experiments}

The ambient-pressure and high-pressure muon-spin rotation/relaxation ($\mu$SR) experiments were conducted at the $\pi$M3 and $\mu$E1 beamlines at the Paul Scherrer Institute (PSI Villigen, Switzerland), using the dedicated GPS (General Purpose Surface, Ref.~\onlinecite{Amato_RSI_2017}) and GPD (General Purpose Decay, Refs.~\onlinecite{Khasanov_HPR_2016, Khasanov_JAP_2022}) muon spectrometers. Quasi-hydrostatic pressures of up to $\simeq 2.3$~GPa were generated using double-wall piston-cylinder clamp cells made of the nonmagnetic MP35N alloy\cite{Khasanov_HPR_2016}. The $\mu$SR data were analyzed using the MUSRFIT software  package\cite{MUSRFIT}.
The $\mu$SR experiments were performed in two modes. In the first mode, zero-field (ZF) $\mu$SR measurements were conducted without applying an external magnetic field. In the second mode, a weak transverse field (WTF) was applied perpendicular to the initial muon-spin polarization.

\subsection{ZF- and WTF-$\mu$SR data analysis procedure}

The experimental $\mu$SR data were fitted using the following functional form:
\begin{equation}
A(t) = A_{\rm 0,s} P_{\rm s}(t) + A_{\rm 0,bg} P_{\rm bg}(t),
\label{eq:asymmetry}
\end{equation}
where the subscripts s and bg denote the sample and background contributions respectively, and $A_{\rm 0}$ and $P$ represent the initial asymmetry and the time evolution of the muon-spin polarization. The sample polarization function, $P_{\rm s}(t)$, was further divided into magnetic (m) and nonmagnetic (nm) components, with corresponding weights $f_{\rm m}$ and $1-f_{\rm m}$. In experiments conducted on the low-background GPS muon spectrometer~\cite{Amato_RSI_2017}, the background contribution was found to be negligible ($A_{\rm 0,bg} \simeq 0$). In $\mu$SR experiments under pressure performed at GPD muon instrument \cite{Khasanov_HPR_2016, Khasanov_JAP_2022}, the background arises from muons stopped in the pressure cell walls and accounts for approximately 50\% of the total $\mu$SR response.

In ZF-$\mu$SR experiments, the magnetic (m) and nonmagnetic (nm) contributions to the sample muon-spin polarization function $P_{\rm s}(t)$ were obtained as:
\begin{equation}
P_{\rm s,m}(t) = \frac{2}{3} \sum_i f_{{\rm m,}i} e^{-\lambda_{{\rm T},i} t} J_0(\gamma_\mu B_{{\rm int},i} t) + \frac{1}{3} e^{-\lambda_{\rm L}},
\label{eq:incommensurate}
\end{equation}
and
\begin{equation}
P_{\rm s,nm}(t) = \frac{2}{3} (1 - \sigma_{\rm GKT}^2 t^2) \exp\left[ -\frac{\sigma_{\rm GKT}^2 t^2}{2} \right] + \frac{1}{3},
\label{eq:GKT}
\end{equation}
where $J_0$ is the zeroth-order Bessel function, $\gamma_\mu = 851.616$~MHz/T is the muon gyromagnetic ratio, $\lambda$ represents the exponential relaxation rates, $\sigma_{\rm GKT}$ is the Gaussian Kubo-Toyabe relaxation rate, and $f_{{\rm m,}i}$s are the volume fractions of individual magnetic components ($\sum f_{{\rm m,}i}=f_{\rm m}$). The coefficients $2/3$ and $1/3$ account for powder averaging, with $2/3$ of the muon spins precessing in internal fields perpendicular (transverse, T) to the field direction and $1/3$ remaining parallel (longitudinal, L) to $B_{{\rm int},i}$\cite{Amato-Morenzoni_book_2024, Schenck_book_1985, Yaouanc_book_2011, Blundell_book_2022}.

The WTF sample response was modeled as:
\begin{eqnarray}
P_{\rm s}(t)&=&f_{\rm m}\left[  \frac{2}{3} e^{-\lambda_{\rm T}t} +\frac{1}{3}  e^{-\lambda_{\rm L}t}  \right] + \nonumber \\
&&(1-f_{\rm m})e^{-\sigma_{\rm WTF}^2t^2/2}\cos(\gamma_\mu B_{\rm WTF} t +\phi),
\label{eq:WTF}
\end{eqnarray}
where $\phi$ is the initial phase of the muon-spin ensemble, $B_{\rm WTF}=5$~mT is the weak transverse field, and $\sigma_{\rm WTF}$ is the Gaussian relaxation rate of the nonmagnetic component. The first term in Eq.~\ref{eq:WTF} corresponds to Eq.~\ref{eq:incommensurate}, where, due to data binning, all oscillating components are combined into a single decay term. The second term describes muon-spin precession in $B_{\rm WTF}$ within the nonmagnetic phase of the sample.

\subsection{Muon stopping sites and dipole-field calculations}

Muon stopping sites were calculated using a DFT$+\mu$ approach~\cite{blundell2023dft} using the MuFinder application~\cite{huddart2022mufinder}.
Density functional theory calculations were performed using the plane-wave pseudopotential code \textsc{castep}~\cite{clark2005first}.
The PBEsol functional~\cite{perdew2008restoring} was used in all calculations.
Calculations were converged to $\sim3$ meV/atom using a plane-wave cutoff of 800 eV and a $3\times3\times3$~$k$-point grid~\cite{monkhorst1976special}.
54 calculations were performed, each where a muon was implanted randomly into a cell of approximately $11\times11\times14$~\AA.
The atomic positions were subsequently allowed to relax until the energy and atomic positions had converged to better than $2\times10^{-2}$~meV/atom and $1\times10^{-3}$~\AA\ respectively, and the maximum force on any atom was less than $5\times10^{-2}$~eV/\AA.
The final muon positions were clustered using the MuFinder application to give the candidate muon stopping sites, which were subsequently used in the \textsc{muesr} code~\cite{bonfa2018introduction} to calculate the dipolar field at these sites for candidate magnetic structures.
To capture the effect of an incommensurate magnetic structure, such as a spin density wave, many calculations with the initial phase of the magnetic structure varied were performed.

\subsection{Resistivity experiments}
Experiments under ambient pressure were performed by using the `Resistivity' option hardware and software of the Quantum Design Physical Property Measurement System (PPMS). Experiments under pressure were performed using the same PPMS instrument by using Almax Easylab Pcell 15/30 module.\cite{Easylab}

\subsection{Neutron powder diffraction experiments}
Neutron powder diffraction (NPD) experiments were carried out using the high-intensity cold neutron diffractometer DMC (neutron wavelength $\lambda_{\rm n}=3.82$~\AA) and the high-resolution thermal neutron diffractometer HRPT ($\lambda_{\rm n}=2.45$~\AA), both located at the spallation neutron source SINQ (PSI Villigen, Switzerland)\cite{DMC, HRPT}. A ca. 9 g sample was loaded into an 8-mm V-container for use in the measurements. The diffraction data were analyzed using the Rietveld refinement program Jana \cite{Jana1, Jana2}.

\subsection{Raman experiments}
Raman spectra were collected using a LabRAM Series Raman Microscope (Horiba Jobin Yvon) with a He–Ne excitation laser ($\lambda = 632.8$~nm, $\sim 20$~mW). The scattered light was dispersed by an 1800 grooves mm$^{-1}$ grating, and an edge filter suppressed the laser line. A 50$\times$ objective was used, with the laser intensity reduced by a factor of 100 to avoid sample heating. Measurements were performed at multiple locations, as the grain size was comparable to the laser spot ($\simeq 50$~$\mu$m), and 20–50 scans were typically averaged to improve statistics. Background subtraction was carried out using the Jobin Yvon software.

\subsection{Other experimental techniques}
%The specific heat measurements, were performed using the PPMS-9 (Physical Property Measurement System) at  the Gda\'{n}sk University of Technology. Measurements were conducted using the `Heat Capacity' option.

Laboratory x-ray powder diffraction was performed at room temperature in the Bragg-Brentano geometry using a Bruker AXS D8 Advance diffractometer (Bruker AXS GmbH, Karlsruhe, Germany) equipped with a Ni-filtered Cu K$\alpha$ radiation and a 1D LynxEye PSD detector to confirm the phase-purity. Le Bail analysis of the obtained diffraction pattern was performed using the FullProf Suite package\cite{fulproof}.

Thermogravimetric Analysis (TGA) hydrogen reduction was performed to elucidate the oxygen content in the resulting La$_4$Ni$_3$O$_{10-\delta}$ samples.
%The initial sample mass was 88.68 mg.
A NETZSCH STA 449F1 Simultaneous Thermal Analyzer (STA) was used for the thermogravimetry experiments. The gas used in the experiment was a mixture of 5 vol.\% of hydrogen (Messer Schweiz AG, 5N) in helium (PanGas, 6N).

Field-cooled magnetization experiments were performed by using SQUID magnetometer Quantum Design MPMS-5 over a wide range of applied fields, from 20~mT to 5~T.

\section*{DATA AVAILABILITY}

All relevant data are available from the authors upon request. The $\mu$SR data are also accessible via the link provided in Ref.~\onlinecite{data-avalibility}.

%\section*{AUTHORS CONTRIBUTIONS}
%R.K. conceived and supervised the project. D.J.G. and I.P. synthesized the sample and conducted x-ray and thermogravimetry characterization. R.K. performed the ambient pressure and high pressure $\mu$SR experiments and analyzed the data. T.J.H. calculated the muon stopping sites and dipole-field distributions for various possible magnetic structures. I.P., L.K., and V.P conducted the neutron powder diffraction experiments and analyzed the data. V.S. and Z.G. conducted electrical transport experiments under pressure with contribution from M.B. S.K., M.J.W., and T.K. conducted the specific heat experiments. R.K. and T.J.H. wrote the manuscript with contributions from I.P, T.K., D.J.G., J.A.K., H.L., and Z.G.

\end{document}